\documentclass[aps,prd,reprint,nofootinbib,superscriptaddress]{revtex4-2}
\usepackage[T1]{fontenc}
\usepackage{lmodern,amsmath,amssymb,bm,microtype,xcolor}
\newenvironment{fullwidth}{\begin{widetext}}{\end{widetext}}
\usepackage[colorlinks=true,citecolor=blue!60!black,linkcolor=blue!60!black]{hyperref}
\newcommand{\Ord}{\mathcal O}
\newcommand{\Hb}{\widehat H}
\newcommand{\Pcal}{\mathcal P}
\newcommand{\be}{\beta_+^2+\beta_-^2}
\begin{document}
\title{Quantum Bianchi IX Anisotropies as Inflationary Relics in the Primordial Spectrum}
\author{Vittorio D'Alelio}
\affiliation{Physics Department, Sapienza University of Rome, P.le A. Moro 5, 00185 Roma, Italy}
\author{Giovanni Montani}
\affiliation{ENEA, C.R. Frascati (Rome), Italy, Via E. Fermi 45, 00044 Frascati (Roma), Italy}
\affiliation{Physics Department, Sapienza University of Rome, P.le A. Moro 5, 00185 Roma, Italy}
\date{\today}
\begin{abstract}
Starting from the Wheeler--DeWitt equation, we investigate whether quantum anisotropies can survive inflation as relics of the preinflationary geometry and leave an imprint on the primordial curvature spectrum. We perform a Born--Oppenheimer separation of the wave functional into slow gravitational and fast matter components, justified by the hierarchy between the Planck scale and the lower matter scale, associated here with the inflaton. The full wave functional is then treated through a semiclassical WKB expansion in $M^{-1}$, where $M$ is quadratic in the Planck mass.

The gravitational sector consists of a de Sitter background supplemented by homogeneous quantum anisotropies in the small-anisotropy regime of the Bianchi IX potential, while matter is represented by a single scalar inflaton field. At order $M$ in the expansion parameter, the Wheeler--DeWitt equation reproduces the vacuum Hamilton--Jacobi equation with a cosmological constant. At order $M^0$, the internal freedom of the Born--Oppenheimer factorization separates quantum field theory on curved spacetime from the anisotropy dynamics, described by time-dependent harmonic oscillators. At order $M^{-1}$, we obtain the quantum-gravitational correction to matter. The anisotropy wave function enters explicitly the corresponding operator, so the corrected dynamics retains information about the Bianchi IX quantum state.

The curvature power spectrum consequently contains two distinguishable contributions: one independent of the anisotropy state and a genuinely anisotropic term. The latter survives through the quantum variance of the anisotropies, thereby identifying the mechanism through which the preinflationary quantum geometry can affect the primordial spectrum.
\end{abstract}
\maketitle

\section{Introduction}
It is well known that a fully consistent and universally accepted proposal for
quantizing the gravitational field is still lacking \cite{ThiemannBook}. At the
same time, a basic question remains open: if gravity is mainly represented by a
classical spacetime metric supplemented by small quantum corrections, can one
derive a consistent quantum field theory on the resulting background
\cite{BirrellDavies,WKBReview}?

Pioneering approaches to the recovery of quantum field theory on curved
spacetime from the full gravity--matter Wheeler--DeWitt equation
\cite{CanonicalMontani} were developed
by Lapchinsky and Rubakov \cite{LapchinskyRubakov} and by Vilenkin
\cite{Vilenkin}. In these formulations, the
standard notion of time evolution is reconstructed from the dependence of the
wave functional on the classical three-metric (for a critical discussion, see
Ref.~\cite{ManicciaAntonini}). The conceptual and dynamical
limitations of this identification have subsequently been analyzed in detail
\cite{DiGioia2021,ManicciaAntonini}. Kiefer and Singh extended Vilenkin's
construction to the next semiclassical order, at which quantum corrections due
to gravity can no longer be neglected
\cite{KieferSingh,KieferBook,BroutVenturi,Bertoni}. The resulting
functional Schr\"odinger equation for matter, however, generally contains
nonunitary contributions \cite{DiGioia2021,WKBReview}.

This difficulty motivated reformulations in which time is supplied by a
material reference system, for instance through a kinematical action or through
the Gaussian reference fluid of Kucha\v r and Torre
\cite{KucharTorre,ManicciaKinematical,Maniccia2022}. Besides preserving a
unitary matter evolution at the relevant order, these constructions emphasize
that the quantum gravitational degrees of freedom cannot simply disappear from
the low-energy theory: they must eventually be averaged over when predictions
are compared with observations
\cite{DiGioia2021,Maniccia2022,ManicciaAntonini,Maniccia2024}. Both the kinematical-action
and Kucha\v r--Torre realizations
\cite{DiGioia2021,Maniccia2022} nevertheless introduce an effective reference
fluid whose energy density is not manifestly positive in the full theory. This
issue is avoided here by adopting Brown--Kucha\v r dust, for which the physical
branch has positive energy density \cite{BrownKuchar}. The clock sector is
treated as part of the fast subsystem, together with ordinary quantum matter,
whereas the gravitational variables constitute the slow sector. This is the
Born--Oppenheimer hierarchy already underlying the original semiclassical
construction \cite{KieferSingh}: it is justified by the large separation
between the characteristic energy scale of matter and the Planck scale of
quantum gravitational dynamics.

One of the most relevant goals of these approaches is to identify possible
corrections to quantum field theory produced by the slowly varying quantum
nature of the gravitational field. A particularly important example is a
modification of the primordial inflationary spectrum
\cite{WeinbergCosmology}. In Ref.~\cite{BrizuelaDeSitter}, the inflationary
spectrum was calculated by neglecting the nonunitary corrections, which were
shown to be small. In the reference-fluid formulation of
Ref.~\cite{Maniccia2024}, instead, the analysis is intrinsically unitary. The
quantum-gravitational correction displays a characteristic dependence on the
scalar-field wavenumber and has a very small amplitude relative to the standard
spectrum, as is naturally expected from the smallness of the perturbative
parameter adopted in the semiclassical expansion.

Here we upgrade the analysis of Ref.~\cite{Maniccia2024} in two fundamental
respects. First, we introduce the physical clock through a genuine
Brown--Kucha\v r dust fluid, whose energy density is positive on the physical
branch. Second, we search for a specific trace in the primordial spectrum of
the quantum-gravity correction associated with the slowly varying graviton
degrees of freedom, which was absent from the original analysis. In the
homogeneous cosmological setting considered here, the graviton contribution is
represented by the two quantum anisotropies of the Bianchi IX geometry: they
are the two genuine cosmological degrees of freedom of the gravitational field,
superposed on the isotropic de Sitter background \cite{MontaniBook}. We
determine the corresponding quantum-gravitational correction to the
inflationary spectrum and show that, once the anisotropy variables are averaged
out of the total wave functional, the result contains a specific contribution
controlled by their quantum state. This anisotropy-dependent contribution can
be distinguished from the part of the correction that is independent of the
anisotropy state.

\subsection{Motivation for the proposed scenario}

Inflation provides a powerful mechanism for the local homogenization and
isotropization of the early Universe \cite{KirillovMontani2002,MontaniBook},
thereby connecting a generic inhomogeneous primordial geometry with the later
Robertson--Walker evolution. Several results support the conjecture that the
preinflationary phase was appropriately described by a geometry substantially
more general than a Robertson--Walker one
\cite{Montani1995,KirillovMontani1997,BeniniMontani2007}. On a given scale, the inflationary process
can nevertheless make the cosmological dynamics indistinguishable from one
driven by homogeneity and isotropy.

For the homogeneous Bianchi IX cosmology considered here, the possibility that
the dynamics naturally approaches an isotropic Universe, supplemented by the
two gravitational degrees of freedom, has long been investigated
\cite{Doroshkevich1973}. In the
inflationary setting, a classical ``bridge'' solution was constructed by
solving the Hamilton--Jacobi equation
\cite{KirillovMontani2002,MontaniBook}. It connects a Kasner epoch of the
Mixmaster dynamics \cite{BKL1970,Misner1968,MisnerMixmaster} with an isotropic inflationary regime: the anisotropy degrees
of freedom are frozen out and effectively vanish because of the large isotropic
contribution associated with the vacuum-energy density driving the inflationary
phase transition.

The present work starts from the idea that, even though the classical
anisotropies are dissipated by the rapid expansion during the de Sitter phase,
their quantum counterparts can survive. It is precisely this relic quantum
component of the Bianchi IX cosmology that we treat as the slowly varying
quantum-gravity correction to the classical isotropic background. In other
words, we study the influence that the quantum anisotropies can induce on the
shape of the inflationary spectrum when the original background cosmology is a
Bianchi IX Universe stretched and isotropized by the vacuum energy.

In this setting, the Brown--Kucha\v r dust \cite{BrownKuchar} has a more direct physical imprint
than the Gaussian fluid used merely to materialize the Kucha\v r--Torre
reference frame. It represents a congruence of pressureless massive particles,
whose proper time defines the relational clock. The microscopic identity of
these particles remains speculative. A possible interpretation is in terms of
stable or sufficiently long-lived particles with masses above the grand
unification scale, already nonrelativistic in the preinflationary epoch. If
$m_{\rm d}$ exceeds the maximum temperature attained during reheating,
$T_{\rm reh}\ll m_{\rm d}$, their thermal production is exponentially
suppressed; they may therefore define the primordial clock without being
regenerated as an appreciable matter component after inflation. This
interpretation is not required by the canonical construction, and possible
nonthermal production channels would have to be assessed in a complete
particle-physics model.

The proposed scenario is thus naturally compatible with a Born--Oppenheimer
separation between a classical isotropic background supplemented by small
quantum anisotropies and the fast matter sector. The latter is represented here
by a single scalar inflaton field, effectively massless during the de Sitter
phase, whose Gaussian fluctuations provide the primordial spectrum analyzed
below.

The paper is organized as follows. In Sec.~II we formulate the
Born--Oppenheimer separation and perform the WKB expansion of the
Wheeler--DeWitt equation through order $M^{-1}$. Section~III solves the quantum
dynamics of the Bianchi IX anisotropies by means of the Lewis--Riesenfeld
invariant. In Sec.~IV we construct the explicit corrected Gaussian matter state,
and in Sec.~V we derive the corresponding primordial power spectrum. Our main
results and their physical interpretation are summarized in Sec.~VI. Technical
details are collected in the appendices.

\section{Born--Oppenheimer expansion}
The canonical and geometrodynamical setting follows the ADM and
Wheeler--DeWitt formulations of general relativity
\cite{ADM,DeWitt,Wheeler,Halliwell,Vilenkin}. We describe the inflationary
minisuperspace by $\alpha=\ln a$ and two small homogeneous Bianchi IX
anisotropies $\bm\beta=(\beta_+,\beta_-)$. This sector is the near-isotropic
limit of the Mixmaster geometry \cite{MisnerMixmaster,MisnerMinisuperspace,RyanShepley}.
Fourier modes $\phi_{\bm k}$ represent free scalar perturbations, and the nearly
constant inflaton potential is absorbed into $\Lambda$. Time is introduced
through the pressureless reference dust of Brown and Kucha\v r
\cite{BrownKuchar,KucharTime,IshamTime}.

\subsection{Brown--Kucha\v r dust as a relational clock}

The Brown--Kucha\v r construction starts from the covariant dust action
\begin{equation}
\begin{aligned}
S_{\rm D}&=-{1\over2}\int d^4x\,\sqrt{-\gamma}\,\varrho
\left(\gamma^{\mu\nu}U_\mu U_\nu+1\right),\\
U_\mu&=-\partial_\mu T+W_j\partial_\mu Z^j .
\end{aligned}
\label{dustaction}
\end{equation}
Here $\varrho$ is the rest-mass density, $Z^j$ are comoving labels for the dust
worldlines, and $T$ is the clock field. Variation with respect to $\varrho$ and
$W_j$ gives $U^\mu U_\mu=-1$ and $U^\mu\partial_\mu Z^j=0$, respectively.
Contracting the definition of $U_\mu$ with $U^\mu$ then yields
$U^\mu\partial_\mu T=1$: hence $T$ measures proper time along each dust
worldline \cite{BrownKuchar}.

The ADM split introduces the canonical pairs $(T,P_T)$ and $(Z^j,P_j)$, with
dust supermomentum
$\mathcal H^{\rm D}_a=P_TT_{,a}+P_jZ^j{}_{,a}$. Using the total
supermomentum constraint, the super-Hamiltonian can be resolved with respect
to $P_T$ as
\begin{equation}
\mathcal C_T=P_T+h=0,\qquad
h=-\sqrt{\mathcal H_\perp^2-q^{ab}\mathcal H_a\mathcal H_b},
\label{dustresolved}
\end{equation}
where $\mathcal H_\perp$ and $\mathcal H_a$ contain only the nondust variables;
the sign follows from the physical branch $P_T>0$. In the homogeneous
comoving reduction, $T_{,a}=\mathcal H_a=0$ and, on the branch
$\mathcal H_\perp<0$, $h=\mathcal H_\perp$. Gauge fixing $Z^j$ therefore leaves
$P_T+H_{\rm phys}=0$, whose quantization gives
$i\hbar\partial_T\Psi=\widehat H_{\rm phys}\Psi$. Evolution is thus measured
by the proper time of the dust rather than extracted from the gravitational
WKB phase.

\subsection{Model setup}

For the present minisuperspace, in conformal gauge $N=e^\alpha$, this gives
\begin{equation}
 (\Hb_{\rm g}+\Hb_\beta+\Hb_\phi+\Hb_{\rm dust})
 \Psi(\alpha,\beta_+,\beta_-;\{\phi_{\bm k}\},T)=0,
\label{wdw}
\end{equation}
with
\begin{align}
\Hb_{\rm g}&={\hbar^2\over48M}e^{-3\alpha}\partial_\alpha^2+4M\Lambda e^{3\alpha},\\
\Hb_\beta&={e^{-3\alpha}\over2}\sum_{\sigma=\pm}[-\hbar^2\partial_{\beta_\sigma}^2+8e^{4\alpha}\beta_\sigma^2],\\
\Hb_\phi&={e^{-\alpha}\over2}\sum_{\bm k}[-\hbar^2\partial_{\phi_{\bm k}}^2+\nu_k^2(\eta)\phi_{\bm k}^2],\\
\Hb_{\rm dust}&=\widehat p_T=-i\hbar{\delta\over\delta T}.
\end{align}
For the effectively massless scalar field on the de Sitter background, the
time-dependent squared frequency is $\nu_k^2(\eta)=k^2-2/\eta^2$.
Our notation and sign conventions for the gravitational sector follow those
of \textit{Gravitation} by Misner, Thorne, and Wheeler
\cite{MisnerThorneWheeler}.
Equivalently, Eq.~\eqref{wdw} can be written compactly as
\begin{equation}
(\widehat p_T+\Hb_{\rm phys})\Psi=0,\qquad
\Hb_{\rm phys}\equiv\Hb_{\rm g}+\Hb_\beta+\Hb_\phi,
\label{deparam}
\end{equation}
The oscillator form of $\Hb_\beta$ follows from the quadratic expansion of the Bianchi IX potential at the isotropic minimum. The expansion scale
\begin{equation}
 M={c^2\over32\pi G}={c\,m_{\rm Pl}^2\over4\hbar}
\label{Mdef}
\end{equation}
is quadratic, not linear, in the Planck mass.
The relation between this expansion in inverse powers of $M$ and the more
conventional semiclassical expansion in powers of $\hbar$ was discussed
explicitly by Maniccia and Montani \cite{Maniccia2022}. For the approximation
considered here the two orderings describe the same semiclassical limit, while
the $M$ expansion has the advantage of isolating the pure-gravity
Hamilton--Jacobi equation at leading order.

The BO separation and WKB expansion are based on the standard molecular
adiabatic construction and its semiclassical extension to quantum gravity
\cite{BornOppenheimer,BroutVenturi,KieferSingh,Bertoni,KieferBook}:
\begin{multline}
\Psi=\psi(\alpha,\bm\beta)\chi(\phi_{\bm k},T;\alpha,\bm\beta)
\\=e^{\frac i\hbar(MS_0+S_1+M^{-1}S_2)}
 e^{\frac i\hbar(Q_1+M^{-1}Q_2)} .
\label{bo}
\end{multline}
We impose
\begin{equation}
{\delta Q_n\over\delta\alpha}=\Ord(M^{-2}),\qquad
{\delta Q_n\over\delta\beta_\sigma}=\Ord(M^{-1}),
\label{adiabatic}
\end{equation}
These two estimates implement the adiabatic hierarchy at two distinct levels.
First, they state that the phase of the fast matter component depends only
parametrically on the gravitational coordinates: matter therefore remains the
fast sector with respect to the geometry. Second, the different powers of
$M$ assigned to the two derivatives introduce a further hierarchy within the
gravitational sector itself. The response along the isotropic direction
$\alpha$ is suppressed more strongly, consistently with the subsequent WKB
reduction of $\alpha$ to a semiclassical de Sitter trajectory. By contrast,
the weaker condition along $\beta_\sigma$ retains the anisotropy dependence at
the next relevant order, so that the variables $\beta_\sigma$ remain genuine
quantum degrees of freedom. The resulting hierarchy is therefore
$\alpha$ (semiclassical background), $\beta_\sigma$ (quantum gravitational
variables), and $\phi_{\bm k}$ (fast matter modes).

Substituting Eq.~\eqref{bo} into the Wheeler--DeWitt equation and applying the two functional estimates in Eq.~\eqref{adiabatic}, the full constraint assumes the hierarchy
\begin{equation}
M\,\mathrm{Eq.}_{1}+\mathrm{Eq.}_{0}+M^{-1}\mathrm{Eq.}_{-1}
+\Ord(M^{-2})=0.
\label{wdwhierarchy}
\end{equation}
Each coefficient must vanish separately.
\subsection{Order \texorpdfstring{$M$}{M}: classical de Sitter background}
At order $M$, the Wheeler--DeWitt equation is
\begin{equation}
\left.\frac{\Hb\Psi}{\Psi}\right|_{M}
=M\left[-{e^{-3\alpha}\over48}S_{0,\alpha}^2
+4\Lambda e^{3\alpha}\right]=0,
\label{wdwM}
\end{equation}
or, equivalently,
\begin{equation}
-{e^{-3\alpha}\over48}(\partial_\alpha S_0)^2+4\Lambda e^{3\alpha}=0.
\label{hj}
\end{equation}
The expanding branch is
\begin{equation}
S_0=-{8\over3}\sqrt{3\Lambda}\,e^{3\alpha},\qquad
S_{0,\alpha}=-8\sqrt{3\Lambda}\,e^{3\alpha}.
\label{s0solution}
\end{equation}
To relate the isotropic variable to conformal time, consider the classical
de Sitter Hamiltonian
\begin{equation}
\mathcal H_{\rm dS}=-{e^{-3\alpha}\over48M}p_\alpha^2
+4M\Lambda e^{3\alpha}.
\end{equation}
Hamilton's equation gives
\begin{equation}
\dot\alpha={\partial\mathcal H_{\rm dS}\over\partial p_\alpha}
=-{e^{-3\alpha}\over24M}p_\alpha,
\end{equation}
and therefore
\begin{equation}
p_\alpha=-24M e^{3\alpha}\dot\alpha.
\end{equation}
Since $MS_0$ is the classical action,
$p_\alpha=M\partial_\alpha S_0$, and hence
\begin{equation}
\partial_\alpha S_0={p_\alpha\over M}
=-24e^{3\alpha}\dot\alpha.
\label{S0momentum}
\end{equation}
It follows directly that
\begin{align}
{\partial S_0\over\partial\alpha}{\partial\over\partial\alpha}
&=-24e^{3\alpha}\dot\alpha\,{\partial\over\partial\alpha}
=-24e^{3\alpha}{d\over dt}\nonumber\\
&=-24e^{3\alpha}e^{-\alpha}{d\over d\eta}
=-24e^{2\alpha}\partial_\eta,
\label{WKBtime}
\end{align}
where $dt=e^\alpha d\eta$ was used in the last line.  Finally, the
second derivative of the Hamilton--Jacobi function is
\begin{equation}
\partial_\alpha^2S_0
=\partial_\alpha\left(-{8\over3}\sqrt{3\Lambda}\,e^{3\alpha}\right)
=-24\sqrt{3\Lambda}\,e^{3\alpha}.
\label{S0second}
\end{equation}
Equations~\eqref{WKBtime} and \eqref{S0second} are the two relations needed
at order $M^0$ to rewrite the gravitational equation as a
conformal-time Schr\"odinger equation for the anisotropies. Thus, the role of
the order-$M$ equation is precisely to determine the semiclassical dynamics of
$\alpha$. At the subsequent order, this dynamics provides the relation between
the conformal time associated with the de Sitter trajectory and the relational
time supplied by the Brown--Kucha\v r dust, allowing the projected
gravitational equation to be recast as the quantum evolution equation for the
anisotropies.

\subsection{Order \texorpdfstring{$M^0$}{M0}: anisotropic quantum gravity and matter}
At order $M^0$, the Wheeler--DeWitt equation reads explicitly
\begin{multline}
0=\left.\frac{\Hb\Psi}{\Psi}\right|_{M^0}
={i\hbar e^{-3\alpha}\over48}S_{0,\alpha\alpha}
-{e^{-3\alpha}\over24}S_{0,\alpha}S_{1,\alpha}\\
+{e^{-3\alpha}\over2}\sum_{\sigma=\pm}
\left[S_{1,\sigma}^2-i\hbar S_{1,\sigma\sigma}
+8e^{4\alpha}\beta_\sigma^2\right]\\
+{e^{-\alpha}\over2}\sum_{\bm k}
\left[Q_{1,\phi_k}^2-i\hbar Q_{1,\phi_k\phi_k}
+\nu_{k,\eta}^2\phi_k^2\right]+Q_{1,T}.
\label{wdwMzero}
\end{multline}
The first two lines define a purely gravitational block
$\mathcal G[S_0,S_1]$, whereas the last line defines the matter--clock block $\mathcal M[Q_1]$. Thus Eq.~\eqref{wdwMzero} is
\begin{equation}
\mathcal G[S_0,S_1]+\mathcal M[Q_1]=0.
\end{equation}
The separation of the total wave functional into slow and fast factors is not
unique, because only their product has a direct physical meaning. In
particular, an arbitrary phase that depends exclusively on the gravitational
configuration variables can be transferred from one factor to the other
without changing $\Psi$. The BO factorization is therefore invariant under
\begin{equation}
\psi_1\longrightarrow e^{i\theta/\hbar}\psi_1,\qquad
\chi_1\longrightarrow e^{-i\theta/\hbar}\chi_1,
\label{BOgauge}
\end{equation}
for any $\theta=\theta(\alpha,\bm\beta)$ independent of the matter modes.
This freedom allows us to choose a BO representation in which the purely
gravitational block satisfies $\mathcal G=0$. Since the total constraint
remains unchanged, Eq.~\eqref{wdwMzero} then requires the matter--clock block
to satisfy $\mathcal M=0$ separately. To display explicitly how conformal time
enters, set
\begin{equation}
\psi_1(\alpha,\bm\beta)=\exp[iS_1(\alpha,\bm\beta)/\hbar].
\end{equation}
Using
\begin{align}
S_{1,\alpha}\psi_1&=-i\hbar\partial_\alpha\psi_1,\nonumber\\
(S_{1,\sigma}^2-i\hbar S_{1,\sigma\sigma})\psi_1
&=-\hbar^2\partial_{\beta_\sigma}^2\psi_1,
\end{align}
the gravitational condition $\mathcal G=0$ becomes
\begin{equation}
{i\hbar e^{-3\alpha}\over24}S_{0,\alpha}
\partial_\alpha\psi_1
+{i\hbar e^{-3\alpha}\over48}S_{0,\alpha\alpha}\psi_1
+\Hb_\beta\psi_1=0.
\label{slowbeforetime}
\end{equation}
Substitution of Eq.~\eqref{WKBtime} and multiplication by $e^\alpha$
then give
\begin{equation}
i\hbar\partial_\eta\psi_1
=e^\alpha\Hb_\beta\psi_1+i\hbar\Gamma(\eta)\psi_1,
\qquad
\Gamma(\eta)={e^{-2\alpha}\over48}S_{0,\alpha\alpha}.
\label{slowtransport}
\end{equation}
Following the reasoning developed by Maniccia, Montani, and Tosoni in
Ref.~\cite{Maniccia2024}, the last term is identified as a
background-dependent WKB transport term. It contains
no anisotropy operator and can be absorbed into a redefinition of the
anisotropy wave function,
\begin{equation}
\bar\psi_1(\eta,\bm\beta)=
\exp\left[-\int_{\eta_i}^{\eta}\Gamma(s)\,ds\right]
\psi_1(\eta,\bm\beta).
\label{transportremove}
\end{equation}
Dropping the bar after this redefinition, the two resulting wave functions are
$\psi_1=\psi_1(\eta,\beta_+,\beta_-)$ and
$\chi_1=\chi_1(\{\phi_{\bm k}\},T;\alpha,\beta_+,\beta_-)$. Their evolution
equations are
\begin{align}
i\hbar\partial_\eta\psi_1&=e^\alpha\Hb_\beta\psi_1,\label{sloweq}\\
i\hbar{\partial\chi_1\over\partial T}&=\Hb_\phi\chi_1.
\label{fasteq}
\end{align}
In the comoving-dust gauge one sets $T=t$, with $t$ the cosmological proper time. Passing to conformal time through ${\rm d}t=a\,{\rm d}\eta$ rewrites the reduced equations in the form used below. Thus time follows from the momentum appearing linearly in Eq.~\eqref{deparam}, not from a WKB derivative constructed out of gravity. Averaging Eq.~\eqref{fasteq} over a normalized anisotropy state then recovers QFT on de Sitter space \cite{BirrellDavies,BunchDavies}.

\subsection{Order \texorpdfstring{$M^{-1}$}{M-1}: quantum-gravity correction}
At total order $M^{-1}$, before the BO gauge fixing, the Wheeler--DeWitt equation is
\begin{multline}
0=\left.\frac{\Hb\Psi}{\Psi}\right|_{M^{-1}}\\
={e^{-3\alpha}\over48M}
\left[i\hbar S_{1,\alpha\alpha}-S_{1,\alpha}^2
-2S_{0,\alpha}S_{2,\alpha}\right]\\
+{e^{-3\alpha}\over2}\sum_{\sigma=\pm}
\left[2S_{1,\sigma}Q_{1,\sigma}-i\hbar Q_{1,\sigma\sigma}\right]\\
+{e^{-3\alpha}\over2M}\sum_{\sigma=\pm}
\left[2S_{1,\sigma}S_{2,\sigma}-i\hbar S_{2,\sigma\sigma}\right]\\
+{e^{-\alpha}\over2M}\sum_{\bm k}
\left[2Q_{1,\phi_k}Q_{2,\phi_k}
-i\hbar Q_{2,\phi_k\phi_k}\right]
+{1\over M}Q_{2,T}.
\label{wdwMinverse}
\end{multline}
The second line has no explicit factor $M^{-1}$ because its order follows from $Q_{1,\sigma}=\Ord(M^{-1})$ and $Q_{1,\sigma\sigma}=\Ord(M^{-1})$. It is precisely this weak dependence on $\beta_\pm$ that must be retained. The order-$M^{-1}$ BO gauge condition removes the terms containing only $S_2$ from the matter equation. Denoting by $\widehat{\mathcal H}_\phi=e^\alpha\Hb_\phi$ the scalar Hamiltonian generating conformal-time evolution, the remaining Wheeler--DeWitt equation becomes
\begin{equation}
i\hbar\partial_\eta\chi_2=\widehat{\mathcal H}_\phi\chi_2
-{e^{-2\alpha}\over2}\sum_{\sigma=\pm}
[\hbar^2\partial_{\beta_\sigma}^2
+2i\hbar(\partial_{\beta_\sigma}S_1)\partial_{\beta_\sigma}]\chi_2,
\label{corrected}
\end{equation}
The displayed anisotropic operator is of relative order $M^{-1}$ by Eq.~\eqref{adiabatic}. It probes the $\bm\beta$ dependence of the corrected matter state and cannot in general be absorbed into a time-dependent phase. Related semiclassical corrections to matter dynamics and inflationary spectra have been investigated in Refs.~\cite{KieferKramer,Kamenshchik,Bini,BrizuelaDeSitter,BrizuelaSlowRoll,Maniccia2022,Maniccia2023,Maniccia2024}.

\section{Time-dependent harmonic oscillators}
We address the anisotropy dynamics using the standard Lewis--Riesenfeld
invariant method for time-dependent quantum harmonic oscillators
\cite{Lewis1967,LewisRiesenfeld,Pedrosa,Dodonov}. Near isotropy the two
$\beta_\sigma$ are identical time-dependent oscillators.
Equation~\eqref{sloweq} has the canonical form
$p_\sigma^2/(2m)+m\kappa^2\beta_\sigma^2/2$, with
\begin{equation}
m(\eta)=a^2(\eta)={1\over H_0^2\eta^2},\qquad \kappa^2=8.
\label{massfrequency}
\end{equation}
For each oscillator the Lewis--Riesenfeld invariant is
\begin{equation}
\widehat I_\sigma={1\over2}\left[
\left({\beta_\sigma\over\rho}\right)^2+
\left(\rho\widehat p_\sigma-m\rho'\beta_\sigma\right)^2\right].
\label{invariant}
\end{equation}
For $\widehat I_\sigma$ to be a dynamical invariant, its total derivative
must vanish:
\begin{equation}
{d\widehat I_\sigma\over d\eta}
\equiv\partial_\eta\widehat I_\sigma
+{1\over i\hbar}[\widehat I_\sigma,\widehat H_\sigma]=0.
\label{invariancecondition}
\end{equation}
Using $[\beta_\sigma,\widehat p_\sigma]=i\hbar$ and
\[
\widehat H_\sigma={\widehat p_\sigma^2\over2m}
+{m\kappa^2\over2}\beta_\sigma^2,
\]
the derivative of Eq.~\eqref{invariant} can be written as
\begin{multline}
{d\widehat I_\sigma\over d\eta}
=-{m\over2}\left[
\rho''+{m'\over m}\rho'+\kappa^2\rho
-{1\over m^2\rho^3}\right]\\
\times\left\{\beta_\sigma,
\rho\widehat p_\sigma-m\rho'\beta_\sigma\right\}.
\label{invariantderivative}
\end{multline}
Since the anticommutator does not vanish identically, condition
\eqref{invariancecondition} is satisfied if and only if $\rho$ obeys the
Ermakov--Pinney equation \cite{Ermakov,Pinney}
\begin{equation}
\rho''+{m'\over m}\rho'+\kappa^2\rho={1\over m^2\rho^3}.
\label{ermakov}
\end{equation}
In our case this becomes
\begin{equation}
\rho''-{2\over\eta}\rho'+\kappa^2\rho
={H_0^4\eta^4\over\rho^3}.
\label{ermakovds}
\end{equation}
Equation~\eqref{ermakovds} admits a family of solutions and therefore does not,
by itself, select a unique quantum state. An additional physical condition is
needed to identify the member of this family corresponding to the vacuum. We
impose this condition in the asymptotic past, where the invariant ground state
is required to coincide with the ground state of the instantaneous
Hamiltonian. The reason why the limit $\eta\to-\infty$ provides the natural
regime for this matching can be seen directly from the oscillator dynamics.
The classical equation
associated with the Hamiltonian defined by Eq.~\eqref{massfrequency} is
\begin{equation}
\beta_\sigma''+{m'\over m}\beta_\sigma'+\kappa^2\beta_\sigma=0.
\end{equation}
Introducing the canonical variable $u_\sigma=\sqrt{m}\,\beta_\sigma$, with
$\sqrt{m}=-(H_0\eta)^{-1}$ on the expanding de Sitter branch, gives
\begin{equation}
u_\sigma''+\Omega_\beta^2(\eta)u_\sigma=0,
\qquad
\Omega_\beta^2(\eta)=\kappa^2-{2\over\eta^2}.
\label{anisotropyfrequency}
\end{equation}
Its relative variation is
\begin{equation}
{\Omega_\beta'\over\Omega_\beta}
={2\over\eta^3\left(\kappa^2-2/\eta^2\right)}
\longrightarrow0
\qquad(\eta\longrightarrow-\infty).
\label{adiabaticpast}
\end{equation}
Equivalently, the standard adiabatic parameter satisfies
$|\Omega_\beta'/\Omega_\beta^2|\rightarrow0$.  Thus, in the remote past the
time scale on which the oscillator frequency varies becomes infinitely long
compared with the oscillation time scale, and positive- and negative-frequency
modes are unambiguously separated.  This is the natural regime in which to
identify the invariant vacuum with the instantaneous Hamiltonian vacuum, in
direct analogy with the Bunch--Davies prescription
\cite{BunchDavies,GrishchukSidorov}.

This matching also supplies the physical criterion needed to remove the
otherwise arbitrary choice among invariant vacua. The invariant vacuum is
preserved by the exact time-dependent evolution, but it does not by itself
specify which state should be regarded as empty of physical excitations. By
contrast, in the asymptotic adiabatic regime the instantaneous Hamiltonian has
a well-defined ground state: it is the state of minimum instantaneous energy
and contains no quanta with respect to the unambiguous positive-frequency
decomposition available there. Requiring the two vacua to coincide therefore
selects the invariant state that approaches this no-particle configuration and
prevents an arbitrary choice of the Ermakov--Pinney constants from introducing
spurious initial excitations. This prescription minimizes particle production
of purely conventional origin; it does not exclude genuine particle production
at later times, when the evolution ceases to be adiabatic.

The corresponding annihilation operators are
\begin{align}
\widehat a_I&={1\over\sqrt{2\hbar}}\left[
{\beta_\sigma\over\rho}
+i(\rho\widehat p_\sigma-m\rho'\beta_\sigma)\right],\nonumber\\
\widehat a_H&=\sqrt{m\kappa\over2\hbar}\,\beta_\sigma
+{i\over\sqrt{2\hbar m\kappa}}\widehat p_\sigma.
\label{annihilationoperators}
\end{align}
Requiring $\widehat a_I$ to approach $\widehat a_H$, up to an irrelevant
phase, gives the two asymptotic conditions
\begin{equation}
m\kappa\rho^2\longrightarrow1,\qquad
m\rho\rho'\longrightarrow0
\qquad(\eta\longrightarrow-\infty).
\label{vacuummatching}
\end{equation}
The first condition matches the Gaussian widths, whereas the second
removes the asymptotic position--momentum correlation.  These conditions
fix the free constants in the Ermakov--Pinney family and select the
positive solution
\begin{equation}
\rho(\eta)={H_0\over\kappa^{3/2}}\sqrt{1+\kappa^2\eta^2}.
\label{rhosolution}
\end{equation}
Indeed, direct substitution verifies Eq.~\eqref{ermakovds}, while
\begin{equation}
m\kappa\rho^2=1+{1\over\kappa^2\eta^2}
\longrightarrow1,\qquad
m\rho\rho'={1\over\kappa\eta}\longrightarrow0.
\label{matchingcheck}
\end{equation}
Thus Eq.~\eqref{rhosolution} is not an arbitrary convenient solution:
it is the member of the Ermakov--Pinney family selected by the physical
vacuum-matching prescription. The limit $\eta\to-\infty$ is retained here as
an asymptotic prescription for selecting this state. Since the quadratic
Bianchi IX approximation independently requires a small anisotropy variance,
the simultaneous validity of the two conditions is not automatic and deserves
further investigation. Appendix~\ref{app:asymptoticconsistency} gives the
corresponding consistency check and identifies the finite early-time window in
which the adiabatic and small-anisotropy requirements can overlap.
To construct a generic state, one must first determine the
Lewis--Riesenfeld phase
\begin{equation}
\Theta=\kappa(\eta-\eta_i)-\arctan(\kappa\eta)+\arctan(\kappa\eta_i).
\label{lrphase}
\end{equation}
Its explicit derivation is given in Appendix~\ref{app:sho}.  After
absorbing the common zero-point phase into the normalization, an
eigenstate of the invariant can be written as
\begin{multline}
\Psi_{n,\sigma}(\beta_\sigma,\eta)
={\mathcal N_\sigma(\eta)e^{-in\Theta}\over\sqrt{2^n n!}}\,
H_n(x_\sigma)\\
\times\exp\left[-{x_\sigma^2\over2}
+{i\over2\hbar}{m\rho'\over\rho}\beta_\sigma^2\right],
\qquad
x_\sigma={\beta_\sigma\over\sqrt{\hbar}\rho}.
\label{invarianteigenstate}
\end{multline}
A generic state is therefore a linear combination
\begin{equation}
\psi_\sigma(\beta_\sigma,\eta)
=\sum_{n=0}^{\infty}c_n\Psi_{n,\sigma}(\beta_\sigma,\eta),
\qquad
\sum_{n=0}^{\infty}|c_n|^2=1.
\label{genericstate}
\end{equation}
It is convenient to collect all the nonuniversal information about the
superposition in
\begin{equation}
F(x,\eta)=\sum_{n=0}^{\infty}
{c_ne^{-in\Theta(\eta,\eta_i)}\over\sqrt{2^n n!}}H_n(x).
\label{Fdefinition}
\end{equation}
For the two independent anisotropies the same construction gives
$F(x_+,x_-,\eta)$, with coefficients $c_{n_+,n_-}$ in the product
Hermite basis.

We now assume that the initial state belongs to the centered, symmetric
Gaussian sector, naturally described in terms of squeezed oscillator states
\cite{Dodonov,GrishchukSidorov}. Besides providing the simplest nontrivial
realization of the generic superposition, this choice is physically supported
by the absence of compelling observational evidence for primordial
non-Gaussianity \cite{PlanckNG}. This is a restriction on the generic
coefficients in Eq.~\eqref{genericstate}, not a consequence of the invariant
method.
The construction and evolution of this Gaussian superposition are
reported in Appendix~\ref{app:gaussianstate}.  The state is characterized
by a squeezing amplitude $r$ and an initial squeezing angle $\vartheta$.
Defining
\begin{equation}
\vartheta_{\rm eff}(\eta)=\vartheta-2\Theta(\eta,\eta_i),
\end{equation}
the corresponding Gaussian state can be written explicitly as
\begin{multline}
\psi_{1,G}(\beta_+,\beta_-,\eta)=\mathcal N_G(\eta)
\exp\left[-{A(\eta)\over2}(x_+^2+x_-^2)\right.\\
\left.+{i\over2\hbar}{m\rho'\over\rho}
(\beta_+^2+\beta_-^2)\right],
\qquad x_\sigma={\beta_\sigma\over\sqrt{\hbar}\rho},
\label{gaussianstate}
\end{multline}
where
\begin{equation}
A(\eta)={1+e^{i\vartheta_{\rm eff}}\tanh r\over
1-e^{i\vartheta_{\rm eff}}\tanh r},
\qquad
|\mathcal N_G|^2={\mathop{\rm Re}A\over\pi\hbar\rho^2}.
\label{gaussianwidth}
\end{equation}
The derivation of Eqs.~\eqref{gaussianstate} and \eqref{gaussianwidth} is
given in Appendix~\ref{app:gaussianstate}.  Taking the modulus squared gives
\begin{align}
|\psi_{1,G}|^2&={\mathop{\rm Re}A\over\pi\hbar\rho^2}
\exp[-\mathop{\rm Re}A(\eta)(x_+^2+x_-^2)],\nonumber\\
\mathop{\rm Re}A&=
{1\over\cosh(2r)-\sinh(2r)\cos\vartheta_{\rm eff}}.
\label{gaussianprobability}
\end{align}
Consequently, for either anisotropy,
\begin{equation}
\langle x_\sigma^2\rangle
={\int_{-\infty}^{\infty}dx_\sigma\,
x_\sigma^2e^{-(\mathop{\rm Re}A)x_\sigma^2}
\over
\int_{-\infty}^{\infty}dx_\sigma\,
e^{-(\mathop{\rm Re}A)x_\sigma^2}}
={1\over2\mathop{\rm Re}A}.
\label{xvariance}
\end{equation}
Since $\beta_\sigma=\sqrt{\hbar}\rho x_\sigma$ and the two directions
are identically distributed, one finally obtains
\begin{equation}
\langle\be\rangle_{r,\vartheta}=\hbar\rho^2
[\cosh(2r)-\sinh(2r)\cos(\vartheta-2\Theta)].
\label{variance}
\end{equation}
For $r=0$ this reduces to $\langle\be\rangle_0=\hbar\rho^2$. Linear
expectation values vanish, while this quadratic quantum contribution
survives.

\section{Explicit solution at order \texorpdfstring{$M^{-1}$}{M-1}}
We now solve explicitly the order-$M^{-1}$ Wheeler--DeWitt equation obtained
in Sec.~II.  To obtain an explicit solution, we adopt a Gaussian ansatz for
the corrected fast component, $\chi_2=\prod_k\chi_k$.  This choice is motivated
both by simplicity and by the absence of compelling observational evidence
for primordial non-Gaussianity \cite{Maldacena,PlanckNG}. For each scalar mode, both the amplitude and
the Gaussian width are allowed to depend on the quantum anisotropies:
\begin{gather}
\chi_k=A_k(\eta,\bm\beta)
e^{-\Omega_k(\eta,\bm\beta)\phi_k^2/2},\\
\Omega_k=\Omega_k^{(0)}+\varepsilon_{\rm QG}\Omega_k^{(1)},
\qquad \varepsilon_{\rm QG}={H_0^2\over M}.
\label{gaussian}
\end{gather}
We now substitute the Gaussian ansatz~\eqref{gaussian} into the corrected
Wheeler--DeWitt equation~\eqref{corrected}. Before linearization, separating
the terms proportional to $\phi_k^0$ and $\phi_k^2$ gives, respectively,
\begin{multline}
i\hbar{\partial_\eta A_k\over A_k}
={\hbar^2\over2}\Omega_k\\
-{e^{-2\alpha}\over2}\sum_\sigma
\left[\hbar^2{A_{k,\sigma\sigma}\over A_k}
+2i\hbar S_{1,\sigma}{A_{k,\sigma}\over A_k}\right],
\label{amplitudeeq}
\end{multline}
\begin{multline}
i\hbar\Omega_k'=\hbar^2\Omega_k^2-\nu_k^2
-e^{-2\alpha}\sum_\sigma\left[
\hbar^2(\partial_{\beta_\sigma}\ln A_k)\Omega_{k,\sigma}\right.\\
\left.+{\hbar^2\over2}\Omega_{k,\sigma\sigma}
+i\hbar S_{1,\sigma}\Omega_{k,\sigma}\right].
\label{exactwidtheq}
\end{multline}
Both $(\partial_{\beta_\sigma}\ln A_k)\Omega_{k,\sigma}$ and
$(\Omega_{k,\sigma})^2$ are $\Ord(\varepsilon_{\rm QG}^2)$. They may be dropped only after deriving these exact equations; the remaining amplitude equation is a consistency condition.
At zeroth order,
\begin{equation}
i\hbar\Omega_k^{(0)\prime}
=\hbar^2(\Omega_k^{(0)})^2-k^2+{2\over\eta^2},
\end{equation}
whose Bunch--Davies solution is
\begin{equation}
\Omega_k^{(0)}={1\over\hbar}\left[
{k^3\eta^2\over1+k^2\eta^2}
+{i\over\eta(1+k^2\eta^2)}\right].
\label{omega0}
\end{equation}
The first-order width obeys
\begin{align}
i\hbar\Omega_k^{(1)\prime}
&=2\hbar^2\Omega_k^{(0)}\Omega_k^{(1)}\nonumber\\
&+e^{-2\alpha}\sum_{\sigma=\pm}\big[-i\hbar(\partial_{\beta_\sigma}S_1)
\partial_{\beta_\sigma}\Omega_k^{(1)}\nonumber\\
&\hspace{12mm}-{\hbar^2\over2}\partial_{\beta_\sigma}^2\Omega_k^{(1)}\big].
\label{omega1eq}
\end{align}
For a regular symmetric Gaussian state write
$\bm\Xi=\nabla_x\ln F=\lambda(\eta)\bm x$, with
$x_\sigma=\beta_\sigma/(\sqrt{\hbar}\rho)$, and define
\begin{equation}
\partial_{\beta_\sigma}S_1=
\left(m{\rho'\over\rho}+{i\over\rho^2}\right)\beta_\sigma
-i{\sqrt{\hbar}\over\rho}\Xi_\sigma ,
\label{s1gradient}
\end{equation}
and hence
\begin{equation}
q_{\rm eff}=m{\rho'\over\rho}+{i\over\rho^2}
-{i\lambda\over\rho^2}.
\label{qeff}
\end{equation}
For simplicity, we assume a quadratic dependence on the anisotropies.  The
corresponding even form of the width at this order is
\begin{equation}
\Omega_k^{(1)}(\eta,\bm\beta)=a_k(\eta)
+b_k(\eta)\left(\be\right)+\Ord(\beta^4).
\label{omegaexp}
\end{equation}
The detailed substitution of this ansatz into Eq.~\eqref{omega1eq}, the
separation of the constant and quadratic parts, and the integration of the
resulting system are reported in Appendix~\ref{app:widthsolution}. To state
the result, define
\begin{align}
\mathcal{F}_k(\eta,\eta_i)&=\exp\!\left[-2i\hbar\int_{\eta_i}^{\eta}
\Omega_k^{(0)}(s){\rm d}s\right],\\
Q(\eta,\eta_i)&=\int_{\eta_i}^{\eta}
e^{-2\alpha(s)}q_{\rm eff}(s){\rm d}s,\\
\tau(\eta,\eta_i)&=\int_{\eta_i}^{\eta}
e^{-2\alpha(s)}e^{-2Q(s,\eta_i)}{\rm d}s .
\end{align}
The explicit solution of Eq.~\eqref{omega1eq} is
\begin{align}
b_k(\eta)&=b_{k,i}\mathcal{F}_k(\eta,\eta_i)e^{-2Q(\eta,\eta_i)},\\
a_k(\eta)&=\mathcal{F}_k(\eta,\eta_i)
[a_{k,i}+2i\hbar b_{k,i}\tau(\eta,\eta_i)],
\end{align}
and hence
\begin{multline}
\Omega_k^{(1)}=\mathcal{F}_k\{a_{k,i}+b_{k,i}
[2i\hbar\tau+e^{-2Q}(\beta_+^2+\beta_-^2)]\}.
\label{omega1solution}
\end{multline}
The homogeneous factor is
\begin{multline}
\mathcal{F}_k={\eta^2(1+k^2\eta_i^2)\over
\eta_i^2(1+k^2\eta^2)}\\
\times e^{-2i[k(\eta-\eta_i)-\arctan(k\eta)+\arctan(k\eta_i)]}.
\label{Fkexact}
\end{multline}
Here $a_{k,i}$ is the isotropic datum and $b_{k,i}$ the initial response to $\be$; their $k$ dependence remains free.

Using Eq.~\eqref{variance}, the averaged correction is
\begin{multline}
\langle\Omega_k^{(1)}\rangle_{r,\vartheta}=\mathcal{F}_k\{a_{k,i}
+b_{k,i}[2i\hbar\tau\\
+e^{-2Q}\hbar\rho^2S_{r,\vartheta}(\eta)]\},
\label{averageomega}\\
S_{r,\vartheta}=\cosh(2r)-\sinh(2r)
\cos[\vartheta-2\Theta(\eta,\eta_i)].
\end{multline}
It is therefore a quantum variance, rather than a classical anisotropic expectation value, that corrects the scalar state.

\section{Primordial power spectrum}
Following the construction of the spectrum from the two-point function used
in Ref.~\cite{Maniccia2024}, we begin with the equal-time correlation function
of the scalar perturbation,
\begin{align}
\Xi(\bm r,\eta)
&=\langle\phi(\eta,\bm x)
\phi(\eta,\bm x+\bm r)\rangle\nonumber\\
&=\int{{\rm d}^3k\over(2\pi)^3}
e^{-i\bm k\cdot\bm r}|f_k(\eta)|^2,
\label{correlation}
\end{align}
where $f_k$ is the Fourier-mode amplitude. The dimensionless spectrum of the
scalar fluctuations is therefore
\begin{equation}
\Pcal_{\phi}(k,\eta)={k^3\over2\pi^2}|f_k(\eta)|^2.
\label{scalarPPS}
\end{equation}
In our BO construction, $|f_k|^2$ must be obtained by first evaluating the
conditional matter variance at fixed $\bm\beta$ and then averaging it over the
normalized quantum state of the anisotropies. Thus
\begin{align}
|f_k|_{\rm eff}^2
&=\int {\rm d}\beta_+\,{\rm d}\beta_-\,
|\psi_1(\bm\beta,\eta)|^2
\langle|\phi_k|^2\rangle_{\bm\beta},
\label{modeaverage}\\
\langle|\phi_k|^2\rangle_{\bm\beta}
&={\int{\rm d}\phi_k\,\phi_k^2|\chi_k|^2
\over\int{\rm d}\phi_k\,|\chi_k|^2}
={1\over2\mathop{\rm Re}\Omega_k}.
\label{variancephi}
\end{align}
The normalization $A_k$ cancels in the conditional ratio, although its
equation remains necessary for consistency. Expanding the denominator before
performing the anisotropy average gives, to first order in
$\varepsilon_{\rm QG}=H_0^2/M$,
\begin{equation}
|f_k|_{\rm eff}^2={1\over2\mathop{\rm Re}\Omega_k^{(0)}}
\left[1-{H_0^2\over M}
{\mathop{\rm Re}\langle\Omega_k^{(1)}\rangle_{r,\vartheta}
\over\mathop{\rm Re}\Omega_k^{(0)}}\right].
\label{modefunctioncorrected}
\end{equation}
The scalar fluctuations can be related to the comoving curvature perturbation
in Fourier space through
\begin{equation}
\mathcal R_{\bm k}=
\sqrt{{4\pi G\over\epsilon}}\,{\phi_{\bm k}\over a},
\label{curvaturefieldrelation}
\end{equation}
Following Ref.~\cite{Maniccia2024}, we compute the mode dynamics in the de Sitter
approximation and retain a small, constant, nonzero $\epsilon$ when converting
scalar fluctuations into curvature perturbations. Consequently, the curvature
perturbation spectrum reads
\begin{equation}
\Pcal_{\mathcal R}(k,\eta)=
{4\pi G\over\epsilon a^2}\Pcal_\phi(k,\eta)
={2Gk^3\over\pi\epsilon a^2}|f_k|_{\rm eff}^2.
\label{curvaturescalarrelation}
\end{equation}
Combining Eqs.~\eqref{scalarPPS}, \eqref{modefunctioncorrected}, and
\eqref{curvaturescalarrelation}, one obtains
\begin{equation}
\Pcal_{\mathcal R}(k,\eta)=\Pcal_{\mathcal R}^{(0)}(k,\eta)
\left[1-{H_0^2\over M}
{\mathop{\rm Re}\langle\Omega_k^{(1)}\rangle_{r,\vartheta}
\over\mathop{\rm Re}\Omega_k^{(0)}}\right].
\label{generalspectrum}
\end{equation}
This quantity must be evaluated in the super-Hubble regime
$k/(aH_0)\ll1$, or equivalently $k\eta\rightarrow0^-$. Once outside the
horizon, the curvature perturbation freezes at the order considered here
\cite{MukhanovFeldmanBrandenberger,Baumann}. In this limit,
\begin{equation}
\mathop{\rm Re}\Omega_k^{(0)}\simeq{k^3\eta^2\over\hbar},
\qquad
\Pcal_{\mathcal R}^{(0)}={\hbar G H_0^2\over\pi\epsilon}.
\end{equation}

Using the super-Hubble form of $\mathop{\rm Re}\Omega_k^{(0)}$ and introducing
a reference scale $k_0$, as conventionally done in the analysis of
quantum-gravitational corrections to the de Sitter spectrum
\cite{BrizuelaDeSitter}, the factor $k^{-3}$ can be displayed explicitly as
$(k_0/k)^3$. To keep this scale dependence manifest, we absorb the reference
normalization of the integration data into
$\bar a_{k,i}\equiv a_{k,i}/k_0^3$ and
$\bar b_{k,i}\equiv b_{k,i}/k_0^3$. Inserting
Eq.~\eqref{averageomega} then gives the central result
\begin{fullwidth}
\begin{equation}
\Pcal_{\mathcal R}(k,\eta)=\Pcal_{\mathcal R}^{(0)}(k,\eta)
\left[1-{H_0^2\over M}\left({k_0\over k}\right)^3
\left(\mathcal C_k^{\rm iso}
+\mathcal C_k^{\rm aniso}S_{r,\vartheta}\right)\right].
\label{spectrum}
\end{equation}
\end{fullwidth}
Here the two dimensionless coefficients are
\begin{align}
\mathcal C_k^{\rm iso}
&={\hbar\over\eta^2}\,
\mathop{\rm Re}\!\left\{\mathcal F_k
\left(\bar a_{k,i}+2i\hbar\tau \bar b_{k,i}\right)\right\},
\label{Ciso}\\
\mathcal C_k^{\rm aniso}
&={\hbar^2\rho^2e^{-2Q}\over\eta^2}\,
\mathop{\rm Re}\!\left\{\mathcal F_k \bar b_{k,i}\right\}.
\label{Caniso}
\end{align}
This decomposition separates the part that is independent of the anisotropy
state from the term proportional to $S_{r,\vartheta}$, which is generated by
the quantum state of the anisotropies.
A centered initial parametrization may be introduced through
\begin{equation}
\Omega_{k,i}^{(1)}=c_{1k}+b_{k,i}
[\beta_+^2+\beta_-^2-\hbar\rho_i^2],
\label{centered}
\end{equation}
so $a_{k,i}=c_{1k}-\hbar\rho_i^2b_{k,i}$. The coefficient $c_{1k}$ is the
isotropic BO datum, whereas $b_{k,i}$ is the anisotropic response. Neither is
fixed by the evolution equation. The explicit $(k_0/k)^3$ factor displays the
quantum-gravitational scale dependence, while $c_{1k}$, $b_{k,i}$ and the
choice $\eta_i(k)$ may introduce further momentum dependence; in particular,
no universal scaling is imposed on $b_{k,i}$.

The anisotropic signature requires a nonvanishing anisotropy variance and a
nonzero matter response $b_{k,i}$ to $\beta_+^2+\beta_-^2$, while the
Born--Oppenheimer hierarchy and $H_0^2/M\ll1$ must remain valid. Consistency
with the small-anisotropy expansion further requires
\begin{equation}
\left\langle\beta_+^2+\beta_-^2\right\rangle
=\hbar\rho^2S_{r,\vartheta}\ll1.
\label{smallanisotropycondition}
\end{equation}
The term proportional to $b_{k,i}\hbar\rho^2S_{r,\vartheta}$ then cannot be
reproduced by $c_{1k}$: it retains information about the width,
Lewis--Riesenfeld phase and squeezing of the preinflationary geometry. Thus
Eq.~\eqref{spectrum} identifies how its quantum state can affect the primordial
spectrum, although a quantitative prediction still requires a physical
prescription for $b_{k,i}(k)$.

Such a prescription may be suggested by the Born--Oppenheimer treatment of
condensed matter, where fast electrons evolve on a slowly varying ionic
background and the electron--phonon interaction determines their
momentum-dependent response \cite{Frohlich,BCS,Mahan}. Here the matter modes
form the fast sector and the Bianchi IX anisotropies the collective slow
gravitational sector. Extending this analogy to a physical interaction kernel
could determine $b_{k,i}(k)$ and make the anisotropic term a testable signature
of the preinflationary geometry.

\section{Conclusions}
We investigated whether a quantum relic of a preinflationary anisotropic
geometry can survive inflation and affect the primordial scalar spectrum,
despite classical isotropization \cite{WaldNoHair}. Starting from the
Wheeler--DeWitt equation, we combined a Born--Oppenheimer separation of slow
gravity and fast matter with a WKB expansion in $M$. The gravitational sector
is a de Sitter background supplemented by two homogeneous quantum Bianchi IX
anisotropies, while Brown--Kucha\v r dust provides physical time.

At order $M$ we recover the vacuum Hamilton--Jacobi equation and the de Sitter
trajectory. At order $M^0$, QFT on this background separates from the two
time-dependent anisotropy oscillators, solved through the Lewis--Riesenfeld
invariant. At order $M^{-1}$ their wave function enters explicitly the
quantum-gravitational operator acting on matter. The corrected Gaussian system
is consistent, since the required nonvanishing amplitude $A_k$ exists, as
shown in Appendix~\ref{app:amplitude}.

Equation~\eqref{spectrum} is our main result. The quantum geometry produces a
departure from the scale-invariant de Sitter spectrum that separates into a
state-independent part, parametrized by $c_{1k}$, and a genuinely anisotropic
term proportional to $b_{k,i}$. The latter depends on the Bianchi IX state
through $\rho$, the Lewis--Riesenfeld phase and $(r,\vartheta)$, and therefore
retains information about the preinflationary geometry. Its momentum profile
is not uniquely fixed here because it also depends on $b_{k,i}(k)$ and on
$\eta_i(k)$; no universal $k^{-3}$ behavior is assumed for this contribution.

Determining that profile requires a more physical description of the
matter--anisotropy interaction. Interpreting the anisotropies as gravitational
analogues of collective lattice modes suggests extending the
Born--Oppenheimer analogy to electron--phonon interactions. This may provide a
mode-dependent prescription for $b_{k,i}$ and establish whether the relic
identified here can become a phenomenologically testable primordial imprint.

\appendix
\section{Detailed solution of the time-dependent anisotropy oscillator}
\label{app:sho}
For completeness, we give the steps leading from the Schr\"odinger equation for the Bianchi IX anisotropies to the states used in the main text.  For either polarization $\sigma=\pm$, Eq.~\eqref{sloweq} can be written as
\begin{equation}
i\hbar\partial_\eta\psi_\sigma=\widehat H_\sigma\psi_\sigma,\qquad
\widehat H_\sigma={\widehat p_\sigma^2\over2m}+{m\kappa^2\over2}\beta_\sigma^2,
\label{app:schobeta}
\end{equation}
where $m=a^2=(H_0^2\eta^2)^{-1}$, $\kappa^2=8$, and $\widehat p_\sigma=-i\hbar\partial_{\beta_\sigma}$.  Consider the quadratic Lewis--Riesenfeld operator
\begin{equation}
\widehat I_\sigma={1\over2}\left[\left({\beta_\sigma\over\rho}\right)^2+
\left(\rho\widehat p_\sigma-m\rho'\beta_\sigma\right)^2\right].
\label{app:invariant}
\end{equation}
Imposing ${\rm d}\widehat I_\sigma/{\rm d}\eta\equiv
\partial_\eta\widehat I_\sigma+(i\hbar)^{-1}[\widehat I_\sigma,\widehat H_\sigma]=0$
and equating the coefficients of $\widehat p_\sigma^2$, $\{\beta_\sigma,\widehat p_\sigma\}$ and $\beta_\sigma^2$ gives a single auxiliary equation,
\begin{equation}
\rho''+{m'\over m}\rho'+\kappa^2\rho={1\over m^2\rho^3}.
\label{app:ep}
\end{equation}
Since $m'/m=-2/\eta$, its de Sitter form is
\begin{equation}
\rho''-{2\over\eta}\rho'+\kappa^2\rho={H_0^4\eta^4\over\rho^3}.
\label{app:epds}
\end{equation}
To verify the positive solution, let $u=1+\kappa^2\eta^2$ and $\rho=C\sqrt u$.  Then
\begin{equation}
\rho'={C\kappa^2\eta\over\sqrt u},\qquad
\rho''={C\kappa^2\over u^{3/2}},
\end{equation}
and the two sides of Eq.~\eqref{app:epds} become, respectively,
\begin{equation}
{C\kappa^6\eta^4\over u^{3/2}},\qquad
{H_0^4\eta^4\over C^3u^{3/2}}.
\end{equation}
Thus $C^4=H_0^4/\kappa^6$, and the positive branch is precisely Eq.~\eqref{rhosolution}.

The Ermakov--Pinney equation alone leaves a family of solutions.  The physical branch is selected by requiring the invariant vacuum to approach the instantaneous Hamiltonian vacuum as $\eta\rightarrow-\infty$.  Their annihilation operators may be chosen as
\begin{align}
\widehat a_I&={1\over\sqrt{2\hbar}}
\left[{\beta_\sigma\over\rho}+i(\rho\widehat p_\sigma-m\rho'\beta_\sigma)\right],\label{app:aI}\\
\widehat a_H&=\sqrt{m\kappa\over2\hbar}\,\beta_\sigma+
{i\over\sqrt{2\hbar m\kappa}}\widehat p_\sigma.\label{app:aH}
\end{align}
Matching their coefficients up to an irrelevant phase yields
\begin{equation}
m\kappa\rho^2\longrightarrow1,\qquad
m\rho\rho'\longrightarrow0.
\label{app:matching}
\end{equation}
For Eq.~\eqref{rhosolution},
\begin{equation}
m\kappa\rho^2=1+{1\over\kappa^2\eta^2},\qquad
m\rho\rho'={1\over\kappa\eta},
\end{equation}
so both requirements hold.  Notice that $\rho'$ itself need not vanish; the physical correlation coefficient $m\rho\rho'$ is the quantity that disappears in the asymptotic past.

The normalized eigenfunctions of $\widehat I_\sigma$ follow by mapping to the unit oscillator coordinate $x_\sigma=\beta_\sigma/(\sqrt\hbar\rho)$:
\begin{multline}
\psi_{n,\sigma}(\beta_\sigma,\eta)={e^{-i(n+1/2)\Theta}\over
\sqrt{2^n n!}(\pi\hbar\rho^2)^{1/4}}
H_n\left({\beta_\sigma\over\sqrt\hbar\rho}\right)\\
\times\exp\left[-{\beta_\sigma^2\over2\hbar\rho^2}
+{i\over2\hbar}{m\rho'\over\rho}\beta_\sigma^2\right].
\label{app:eigenstates}
\end{multline}
The Lewis--Riesenfeld phase is fixed by
\begin{equation}
\Theta(\eta,\eta_i)=\int_{\eta_i}^{\eta}{ds\over m(s)\rho^2(s)}.
\end{equation}
Using Eq.~\eqref{rhosolution}, one obtains
\begin{equation}
\Theta=\kappa(\eta-\eta_i)-\arctan(\kappa\eta)
+\arctan(\kappa\eta_i),
\label{app:phase}
\end{equation}
whose derivative is $1/(m\rho^2)$ as required.  Equations~\eqref{app:eigenstates} and \eqref{app:phase} therefore provide the exact time-dependent oscillator solution employed in Sec.~III.

\subsection{Construction of the Gaussian state}
\label{app:gaussianstate}
For two anisotropies the most general pure state in the invariant basis is
\begin{align}
|\psi_1\rangle&=\sum_{n_+,n_-=0}^{\infty}
c_{n_+,n_-}|n_+,n_-\rangle_I,\nonumber\\
\sum_{n_+,n_-}|c_{n_+,n_-}|^2&=1.
\end{align}
Introducing $x_\sigma=\beta_\sigma/(\sqrt{\hbar}\rho)$, its coordinate
representation can be written as a universal Gaussian factor multiplying
\begin{multline}
F(x_+,x_-,\eta)=
\sum_{n_+,n_-=0}^{\infty}
{c_{n_+,n_-}e^{-i(n_++n_-)\Theta}\over
\sqrt{2^{n_++n_-}n_+!n_-!}}\\
\times H_{n_+}(x_+)H_{n_-}(x_-),
\label{app:Fgeneral}
\end{multline}
where the common zero-point phase has been absorbed into the normalization.

We restrict the initial data to the regular, centered, and symmetric
Gaussian sector.  Equivalently, we impose
\begin{equation}
\bm\nabla_x\ln F=\lambda(\eta)\bm x.
\label{app:gaussianassumption}
\end{equation}
Integrating separately with respect to $x_+$ and $x_-$ gives
\begin{equation}
F_G(x_+,x_-,\eta)=F_0(\eta)
\exp\left[{\lambda(\eta)\over2}(x_+^2+x_-^2)\right].
\label{app:FG}
\end{equation}
After combining this factor with the universal Gaussian, define
$A(\eta)=1-\lambda(\eta)$.  Normalizability requires
$\mathop{\rm Re}A>0$, and the wave function takes the form
\begin{multline}
\psi_{1,G}=\mathcal N_G(\eta)
\exp\left[-{A(\eta)\over2}(x_+^2+x_-^2)\right.\\
\left.+{i\over2\hbar}{m\rho'\over\rho}
(\beta_+^2+\beta_-^2)\right].
\label{app:gaussianwavefunction}
\end{multline}

The time dependence of $\lambda$ is fixed by the Schr\"odinger equation.
Equating the quadratic coefficients yields the Riccati equation
\begin{equation}
{d\lambda\over d\Theta}=i\lambda(\lambda-2).
\label{app:riccati}
\end{equation}
The fractional transformation
\begin{equation}
z={\lambda\over2-\lambda},\qquad
\lambda={2z\over1+z},
\end{equation}
linearizes it:
\begin{equation}
{dz\over d\Theta}=-2iz,\qquad
z(\eta)=z_i e^{-2i\Theta(\eta,\eta_i)}.
\end{equation}
Parametrize the initial Gaussian through the squeezing data
\begin{equation}
z_i=-e^{i\vartheta}\tanh r,\qquad r\geq0.
\end{equation}
With $\vartheta_{\rm eff}=\vartheta-2\Theta$, one finds
\begin{equation}
A={1-z\over1+z}
={1+e^{i\vartheta_{\rm eff}}\tanh r\over
1-e^{i\vartheta_{\rm eff}}\tanh r},
\end{equation}
and therefore
\begin{equation}
\mathop{\rm Re}A=
{1\over\cosh(2r)-\sinh(2r)\cos\vartheta_{\rm eff}}.
\label{app:ReA}
\end{equation}
For each direction, the elementary Gaussian integral gives
\begin{equation}
\langle x_\sigma^2\rangle={1\over2\mathop{\rm Re}A}.
\end{equation}
Using $\beta_\sigma=\sqrt{\hbar}\rho x_\sigma$ and summing over
$\sigma=\pm$ reproduces Eq.~\eqref{variance}.

\section{Order-$M^{-1}$ Gaussian dynamics}
\label{app:orderMinverse}

\subsection{\texorpdfstring{Detailed solution of Eq.~\eqref{omega1eq}}{Detailed solution of the first-order width equation}}
\label{app:widthsolution}
We derive here the solution for the first-order Gaussian width used in
Sec.~IV. For the centered symmetric anisotropy state,
Eq.~\eqref{s1gradient} reduces to
\begin{equation}
\partial_{\beta_\sigma}S_1=q_{\rm eff}(\eta)\beta_\sigma,
\qquad
q_{\rm eff}=m{\rho'\over\rho}+{i(1-\lambda)\over\rho^2}.
\label{app:qeff}
\end{equation}
With $R_\beta^2=\beta_+^2+\beta_-^2$, insert the quadratic ansatz
\begin{equation}
\Omega_k^{(1)}(\eta,\bm\beta)=a_k(\eta)+b_k(\eta)R_\beta^2
\label{app:widthansatz}
\end{equation}
into Eq.~\eqref{omega1eq}. Its anisotropy derivatives are
\begin{align}
\partial_{\beta_\sigma}\Omega_k^{(1)}&=2b_k\beta_\sigma,
&\partial_{\beta_\sigma}^2\Omega_k^{(1)}&=2b_k,
\end{align}
and therefore
\begin{align}
\sum_{\sigma=\pm}(\partial_{\beta_\sigma}S_1)
\partial_{\beta_\sigma}\Omega_k^{(1)}
&=2q_{\rm eff}b_kR_\beta^2,
\label{app:driftterm}\\
\sum_{\sigma=\pm}\partial_{\beta_\sigma}^2\Omega_k^{(1)}
&=4b_k.
\label{app:laplacianterm}
\end{align}
The constant and quadratic powers of $R_\beta$ must vanish independently,
which gives the triangular system
\begin{align}
i\hbar a_k'&=2\hbar^2\Omega_k^{(0)}a_k
-2\hbar^2e^{-2\alpha}b_k,
\label{app:aeq}\\
i\hbar b_k'&=2\hbar^2\Omega_k^{(0)}b_k
-2i\hbar e^{-2\alpha}q_{\rm eff}b_k.
\label{app:beq}
\end{align}
The second equation is autonomous. Introducing
\begin{align}
\mathcal F_k(\eta,\eta_i)
&=\exp\left[-2i\hbar\int_{\eta_i}^{\eta}
\Omega_k^{(0)}(s){\rm d}s\right],
\label{app:Fdef}\\
Q(\eta,\eta_i)&=\int_{\eta_i}^{\eta}
e^{-2\alpha(s)}q_{\rm eff}(s){\rm d}s,
\label{app:Qdef}
\end{align}
its solution with $b_k(\eta_i)=b_{k,i}$ is
\begin{equation}
b_k(\eta)=b_{k,i}\mathcal F_k(\eta,\eta_i)
e^{-2Q(\eta,\eta_i)}.
\label{app:bsolution}
\end{equation}
To solve Eq.~\eqref{app:aeq}, write
$a_k=\mathcal F_k y_k$. The homogeneous contribution cancels and one obtains
\begin{equation}
y_k'=2i\hbar b_{k,i}e^{-2\alpha}e^{-2Q}.
\end{equation}
Defining
\begin{equation}
\tau(\eta,\eta_i)=\int_{\eta_i}^{\eta}
e^{-2\alpha(s)}e^{-2Q(s,\eta_i)}{\rm d}s,
\label{app:taudef}
\end{equation}
and imposing $a_k(\eta_i)=a_{k,i}$ gives
\begin{equation}
a_k(\eta)=\mathcal F_k(\eta,\eta_i)
\left[a_{k,i}+2i\hbar b_{k,i}\tau(\eta,\eta_i)\right].
\label{app:asolution}
\end{equation}
Combining Eqs.~\eqref{app:widthansatz}, \eqref{app:bsolution}, and
\eqref{app:asolution} reproduces Eq.~\eqref{omega1solution}.

\subsection{Consistency of the Gaussian solution}
\label{app:amplitude}
We now prove explicitly that the corrected Gaussian width admits a compatible amplitude.  For each matter mode we retain the full anisotropy dependence,
\begin{equation}
\chi_k=A_k(\eta,\bm\beta)
\exp[-\Omega_k(\eta,\bm\beta)\phi_k^2/2].
\label{app:gaussian}
\end{equation}
Writing $E_k=\exp(-\Omega_k\phi_k^2/2)$ gives
\begin{align}
\partial_{\beta_\sigma}(A_kE_k)&=(A_{k,\sigma}-{1\over2}A_k\Omega_{k,\sigma}\phi_k^2)E_k,\nonumber\\
\partial_{\beta_\sigma}^2(A_kE_k)&=\left[A_{k,\sigma\sigma}
-A_{k,\sigma}\Omega_{k,\sigma}\phi_k^2\right.\nonumber\\
&\left.\quad-{1\over2}A_k\Omega_{k,\sigma\sigma}\phi_k^2
+{1\over4}A_k\Omega_{k,\sigma}^2\phi_k^4\right]E_k.
\label{app:betaderivatives}
\end{align}
Moreover,
\begin{equation}
\widehat{\mathcal H}_{\phi,k}\chi_k=
\left[{\hbar^2\over2}\Omega_k+{1\over2}
(\nu_k^2-\hbar^2\Omega_k^2)\phi_k^2\right]\chi_k.
\end{equation}
Inserting these relations into Eq.~\eqref{corrected} and separating the coefficients of $\phi_k^0$ and $\phi_k^2$ yields
\begin{multline}
i\hbar{\partial_\eta A_k\over A_k}={\hbar^2\over2}\Omega_k
-{e^{-2\alpha}\over2}\sum_\sigma\left[
\hbar^2{A_{k,\sigma\sigma}\over A_k}\right.\\
\left.+2i\hbar S_{1,\sigma}{A_{k,\sigma}\over A_k}\right],
\label{app:amplitudeeq}
\end{multline}
\begin{multline}
i\hbar\Omega_k'=\hbar^2\Omega_k^2-\nu_k^2
-e^{-2\alpha}\sum_\sigma\left[
\hbar^2(\partial_{\beta_\sigma}\ln A_k)\Omega_{k,\sigma}\right.\\
\left.+{\hbar^2\over2}\Omega_{k,\sigma\sigma}
+i\hbar S_{1,\sigma}\Omega_{k,\sigma}\right].
\label{app:widtheq}
\end{multline}
The residual coefficient of $\phi_k^4$ is proportional to
$e^{-2\alpha}\hbar^2\sum_\sigma\Omega_{k,\sigma}^2$.  Since
$\Omega_{k,\sigma}=\Ord(M^{-1})$, this term is $\Ord(M^{-2})$ and is consistently absent at the order considered.  Likewise,
$(\partial_{\beta_\sigma}\ln A_k)\Omega_{k,\sigma}=\Ord(M^{-2})$ once the perturbative solution below is used.  These terms are discarded only after deriving Eqs.~\eqref{app:amplitudeeq} and \eqref{app:widtheq}.

Let $\varepsilon=M^{-1}$ and write
\begin{equation}
\Omega_k=\Omega_k^{(0)}+\varepsilon\Omega_k^{(1)}+\Ord(\varepsilon^2),\qquad
A_k=A_k^{(0)}(\eta)e^{\varepsilon\mathcal F_k(\eta,\bm\beta)}.
\label{app:expansions}
\end{equation}
This parametrization guarantees $A_k\neq0$.  The zeroth-order part of Eq.~\eqref{app:amplitudeeq} is
\begin{equation}
i\hbar{(A_k^{(0)})'\over A_k^{(0)}}={\hbar^2\over2}\Omega_k^{(0)},
\end{equation}
and hence
\begin{equation}
A_k^{(0)}(\eta)=A_{k,0}
\exp\left[-{i\hbar\over2}\int_{\eta_*}^{\eta}
\Omega_k^{(0)}(s)\,ds\right].
\label{app:A0}
\end{equation}
At first order the amplitude equation becomes the linear PDE
\begin{equation}
i\hbar\mathcal F_k'={\hbar^2\over2}\Omega_k^{(1)}
+e^{-2\alpha}\sum_\sigma\left[-i\hbar S_{1,\sigma}\mathcal F_{k,\sigma}
-{\hbar^2\over2}\mathcal F_{k,\sigma\sigma}\right].
\label{app:FPDE}
\end{equation}

The width found in Sec.~IV is quadratic,
$\Omega_k^{(1)}=a_k+b_kR_\beta^2$, where
$R_\beta^2=\beta_+^2+\beta_-^2$.  For the centered anisotropy state one may write
\begin{equation}
S_{1,\sigma}=\Gamma(\eta)\beta_\sigma,\qquad
\Gamma={m\rho'\over\rho}+{i(1-\lambda)\over\rho^2}
\equiv q_{\rm eff}.
\label{app:Gamma}
\end{equation}
Because the source lies in $\mathrm{span}\{1,R_\beta^2\}$, take
\begin{equation}
\mathcal F_k=f_k(\eta)+g_k(\eta)R_\beta^2.
\label{app:Fquadratic}
\end{equation}
Then
\begin{equation}
\sum_\sigma S_{1,\sigma}\mathcal F_{k,\sigma}=2\Gamma g_kR_\beta^2,\qquad
\sum_\sigma\mathcal F_{k,\sigma\sigma}=4g_k.
\end{equation}
No quartic power is generated: the two-dimensional polynomial space is closed under the differential operator in Eq.~\eqref{app:FPDE}.  Equating its constant and quadratic parts gives
\begin{align}
i\hbar f_k'&={\hbar^2\over2}a_k-2\hbar^2e^{-2\alpha}g_k,
\label{app:feq}\\
i\hbar g_k'&={\hbar^2\over2}b_k-2i\hbar e^{-2\alpha}\Gamma g_k.
\label{app:geq}
\end{align}
Define the integrating factor
\begin{equation}
I_\Gamma(\eta)=\exp\left[2\int_{\eta_*}^{\eta}
e^{-2\alpha(s)}\Gamma(s)\,ds\right].
\end{equation}
Equations~\eqref{app:feq} and \eqref{app:geq} are solved by
\begin{align}
g_k(\eta)&=I_\Gamma^{-1}(\eta)
\left[g_{k,*}-{i\hbar\over2}\int_{\eta_*}^{\eta}
I_\Gamma(s)b_k(s)\,ds\right],
\label{app:gsolution}\\
f_k(\eta)&=f_{k,*}-{i\hbar\over2}
\int_{\eta_*}^{\eta}a_k(s)\,ds\nonumber\\
&\quad+2i\hbar\int_{\eta_*}^{\eta}
e^{-2\alpha(s)}g_k(s)\,ds.
\label{app:fsolution}
\end{align}
Combining Eqs.~\eqref{app:A0}, \eqref{app:Fquadratic}, \eqref{app:gsolution}, and \eqref{app:fsolution}, one obtains the explicit nonvanishing amplitude
\begin{multline}
A_k(\eta,\bm\beta)=A_{k,0}
\exp\left[-{i\hbar\over2}\int_{\eta_*}^{\eta}
\Omega_k^{(0)}(s)\,ds\right]\\
\times\exp\left\{{1\over M}
[f_k(\eta)+g_k(\eta)R_\beta^2]+\Ord(M^{-2})\right\}.
\label{app:Aexplicit}
\end{multline}
For de Sitter space, $e^{-2\alpha}=H_0^2\eta^2$ and
$e^{-2\alpha}\Gamma=\rho'/\rho+i e^{-2\alpha}(1-\lambda)/\rho^2$,
which is regular for $\eta<0$ whenever $\lambda(\eta)$ is regular.  Hence $I_\Gamma$, $f_k$, and $g_k$ exist throughout the inflationary interval.  Finally, conditional normalization at fixed $\bm\beta$ fixes the modulus through
\begin{equation}
|A_k|^2=\sqrt{{\mathop{\rm Re}\Omega_k\over\pi}},
\qquad
2\mathop{\rm Re}\mathcal F_k={1\over2}
{\mathop{\rm Re}\Omega_k^{(1)}\over
\mathop{\rm Re}\Omega_k^{(0)}}.
\label{app:normalization}
\end{equation}
The phase is simultaneously fixed by the PDE.  This proves constructively that the width and amplitude equations are mutually consistent through $\Ord(M^{-1})$.

\section{Consistency of the asymptotic vacuum prescription}
\label{app:asymptoticconsistency}
The vacuum prescription used in Sec.~III is imposed through the asymptotic
behavior of the solution as $\eta\to-\infty$. This limit guarantees
adiabaticity, but the harmonic approximation to the Bianchi IX potential also
requires the anisotropies to remain small. We therefore compare the two
conditions explicitly.

For the canonical anisotropy mode, Eq.~\eqref{anisotropyfrequency} gives
\begin{equation}
\Omega_\beta(\eta)=\sqrt{\kappa^2-{2\over\eta^2}},
\qquad \kappa^2\eta^2>2.
\end{equation}
The standard adiabatic parameter is
\begin{equation}
\mathcal A(\eta)
\equiv\left|{\Omega_\beta'\over\Omega_\beta^2}\right|
={2\over |\eta|^3\left(\kappa^2-2/\eta^2\right)^{3/2}}.
\label{app:adiabaticparameter}
\end{equation}
Introducing $x\equiv\kappa|\eta|$, this becomes
\begin{equation}
\mathcal A(x)={2\over(x^2-2)^{3/2}},
\label{app:adiabaticx}
\end{equation}
and hence $\mathcal A\ll1$ for $x\gg1$, consistently with the asymptotic
matching adopted in Sec.~III.

On the other hand, Eqs.~\eqref{rhosolution} and \eqref{variance} give
\begin{equation}
\left\langle\beta_+^2+\beta_-^2\right\rangle
={\hbar H_0^2\over\kappa^3}(1+x^2)S_{r,\vartheta}.
\label{app:anisotropyvariance}
\end{equation}
Since $S_{r,\vartheta}(\eta)$ oscillates between $e^{-2|r|}$ and
$e^{2|r|}$, a state-independent sufficient bound over the full phase evolution
is obtained by defining
\begin{equation}
\delta_\beta^{\rm max}\equiv
{\hbar H_0^2\over\kappa^3}e^{2|r|}.
\end{equation}
The near-isotropic expansion requires
\begin{equation}
\delta_\beta^{\rm max}(1+x^2)\ll1.
\label{app:smallcondition}
\end{equation}
Thus the strict limit $x\to\infty$ cannot itself belong to the
small-anisotropy domain for fixed nonzero $\delta_\beta^{\rm max}$. The asymptotic limit
must therefore be understood as the mathematical condition selecting the
Ermakov--Pinney branch, rather than as a physical epoch at which the quadratic
Bianchi IX approximation is asserted to hold arbitrarily far into the past.

The relevant consistency question is whether a finite early-time interval
exists in which the selected state is already adiabatic while its anisotropy
variance remains perturbatively small. Parametrically, the two requirements
are
\begin{equation}
1\ll x\ll{1\over\sqrt{\delta_\beta^{\rm max}}}
=\left({\kappa^3\over\hbar H_0^2e^{2|r|}}\right)^{1/2}.
\label{app:overlapwindow}
\end{equation}
Such an overlap exists provided $\delta_\beta^{\rm max}\ll1$. More precisely, for chosen
tolerances $\mathcal A<\varepsilon_{\rm ad}$ and
$\langle\beta_+^2+\beta_-^2\rangle<\varepsilon_\beta$, one must choose $x_i$
such that
\begin{equation}
\sqrt{2+\left({2\over\varepsilon_{\rm ad}}\right)^{2/3}}
<x_i<
\sqrt{{\varepsilon_\beta\over\delta_\beta^{\rm max}}-1}.
\label{app:finitewindow}
\end{equation}
Equation~\eqref{app:finitewindow} provides a direct consistency test for any
choice of $H_0$, squeezing parameters and initial time. In the present work we
retain the asymptotic prescription because it uniquely fixes the oscillator
state, while restricting the physical use of that solution to an overlap
window of the form \eqref{app:overlapwindow}. Establishing this window for a
fully specified preinflationary completion, and determining whether the
quadratic Bianchi IX regime can be extended beyond it, are left for future
work.

\scriptsize
\bibliographystyle{apsrev4-2}
\bibliography{paperDalelioMontani}

\begin{thebibliography}{59}%
\makeatletter
\providecommand \@ifxundefined [1]{%
 \@ifx{#1\undefined}
}%
\providecommand \@ifnum [1]{%
 \ifnum #1\expandafter \@firstoftwo
 \else \expandafter \@secondoftwo
 \fi
}%
\providecommand \@ifx [1]{%
 \ifx #1\expandafter \@firstoftwo
 \else \expandafter \@secondoftwo
 \fi
}%
\providecommand \natexlab [1]{#1}%
\providecommand \enquote  [1]{``#1''}%
\providecommand \bibnamefont  [1]{#1}%
\providecommand \bibfnamefont [1]{#1}%
\providecommand \citenamefont [1]{#1}%
\providecommand \href@noop [0]{\@secondoftwo}%
\providecommand \href [0]{\begingroup \@sanitize@url \@href}%
\providecommand \@href[1]{\@@startlink{#1}\@@href}%
\providecommand \@@href[1]{\endgroup#1\@@endlink}%
\providecommand \@sanitize@url [0]{\catcode `\\12\catcode `\$12\catcode
  `\&12\catcode `\#12\catcode `\^12\catcode `\_12\catcode `\%12\relax}%
\providecommand \@@startlink[1]{}%
\providecommand \@@endlink[0]{}%
\providecommand \url  [0]{\begingroup\@sanitize@url \@url }%
\providecommand \@url [1]{\endgroup\@href {#1}{\urlprefix }}%
\providecommand \urlprefix  [0]{URL }%
\providecommand \Eprint [0]{\href }%
\providecommand \doibase [0]{https://doi.org/}%
\providecommand \selectlanguage [0]{\@gobble}%
\providecommand \bibinfo  [0]{\@secondoftwo}%
\providecommand \bibfield  [0]{\@secondoftwo}%
\providecommand \translation [1]{[#1]}%
\providecommand \BibitemOpen [0]{}%
\providecommand \bibitemStop [0]{}%
\providecommand \bibitemNoStop [0]{.\EOS\space}%
\providecommand \EOS [0]{\spacefactor3000\relax}%
\providecommand \BibitemShut  [1]{\csname bibitem#1\endcsname}%
\let\auto@bib@innerbib\@empty
\bibitem [{\citenamefont {Thiemann}(2007)}]{ThiemannBook}%
  \BibitemOpen
  \bibfield  {author} {\bibinfo {author} {\bibfnamefont {T.}~\bibnamefont
  {Thiemann}},\ }\href@noop {} {}\bibinfo {howpublished} {\textit{Modern
  Canonical Quantum General Relativity} (Cambridge University Press,
  Cambridge)} (\bibinfo {year} {2007})\BibitemShut {NoStop}%
\bibitem [{\citenamefont {Birrell}\ and\ \citenamefont
  {Davies}(1982)}]{BirrellDavies}%
  \BibitemOpen
  \bibfield  {author} {\bibinfo {author} {\bibfnamefont {N.~D.}\ \bibnamefont
  {Birrell}}\ and\ \bibinfo {author} {\bibfnamefont {P.~C.~W.}\ \bibnamefont
  {Davies}},\ }\href@noop {} {}\bibinfo {howpublished} {\textit{Quantum Fields
  in Curved Space} (Cambridge University Press, Cambridge)} (\bibinfo {year}
  {1982})\BibitemShut {NoStop}%
\bibitem [{\citenamefont {Maniccia}\ \emph {et~al.}(2022)\citenamefont
  {Maniccia}, \citenamefont {Angelis},\ and\ \citenamefont
  {Montani}}]{WKBReview}%
  \BibitemOpen
  \bibfield  {author} {\bibinfo {author} {\bibfnamefont {G.}~\bibnamefont
  {Maniccia}}, \bibinfo {author} {\bibfnamefont {M.~D.}\ \bibnamefont
  {Angelis}},\ and\ \bibinfo {author} {\bibfnamefont {G.}~\bibnamefont
  {Montani}},\ }\href@noop {} {}\bibinfo {howpublished} {Universe \textbf{8},
  556} (\bibinfo {year} {2022})\BibitemShut {NoStop}%
\bibitem [{\citenamefont {Cianfrani}\ \emph {et~al.}(2014)\citenamefont
  {Cianfrani}, \citenamefont {Lecian}, \citenamefont {Lulli},\ and\
  \citenamefont {Montani}}]{CanonicalMontani}%
  \BibitemOpen
  \bibfield  {author} {\bibinfo {author} {\bibfnamefont {F.}~\bibnamefont
  {Cianfrani}}, \bibinfo {author} {\bibfnamefont {O.~M.}\ \bibnamefont
  {Lecian}}, \bibinfo {author} {\bibfnamefont {M.}~\bibnamefont {Lulli}},\ and\
  \bibinfo {author} {\bibfnamefont {G.}~\bibnamefont {Montani}},\ }\href@noop
  {} {}\bibinfo {howpublished} {\textit{Canonical Quantum Gravity: Fundamentals
  and Recent Developments} (World Scientific, Singapore), doi:10.1142/8957}
  (\bibinfo {year} {2014})\BibitemShut {NoStop}%
\bibitem [{\citenamefont {Lapchinsky}\ and\ \citenamefont
  {Rubakov}(1979)}]{LapchinskyRubakov}%
  \BibitemOpen
  \bibfield  {author} {\bibinfo {author} {\bibfnamefont {V.~G.}\ \bibnamefont
  {Lapchinsky}}\ and\ \bibinfo {author} {\bibfnamefont {V.~A.}\ \bibnamefont
  {Rubakov}},\ }\href@noop {} {}\bibinfo {howpublished} {Canonical quantization
  of gravity and quantum field theory in curved space-time, Acta Phys. Pol. B
  \textbf{10}, 1041--1048} (\bibinfo {year} {1979})\BibitemShut {NoStop}%
\bibitem [{\citenamefont {Vilenkin}(1989)}]{Vilenkin}%
  \BibitemOpen
  \bibfield  {author} {\bibinfo {author} {\bibfnamefont {A.}~\bibnamefont
  {Vilenkin}},\ }\href@noop {} {}\bibinfo {howpublished} {Phys. Rev. D
  \textbf{39}, 1116} (\bibinfo {year} {1989})\BibitemShut {NoStop}%
\bibitem [{\citenamefont {Maniccia}\ \emph
  {et~al.}(2023{\natexlab{a}})\citenamefont {Maniccia}, \citenamefont
  {Montani},\ and\ \citenamefont {Antonini}}]{ManicciaAntonini}%
  \BibitemOpen
  \bibfield  {author} {\bibinfo {author} {\bibfnamefont {G.}~\bibnamefont
  {Maniccia}}, \bibinfo {author} {\bibfnamefont {G.}~\bibnamefont {Montani}},\
  and\ \bibinfo {author} {\bibfnamefont {S.}~\bibnamefont {Antonini}},\
  }\href@noop {} {}\bibinfo {howpublished} {Phys. Rev. D \textbf{107}, L061901}
  (\bibinfo {year} {2023}{\natexlab{a}})\BibitemShut {NoStop}%
\bibitem [{\citenamefont {Gioia}\ \emph {et~al.}(2021)\citenamefont {Gioia},
  \citenamefont {Maniccia}, \citenamefont {Montani},\ and\ \citenamefont
  {Niedda}}]{DiGioia2021}%
  \BibitemOpen
  \bibfield  {author} {\bibinfo {author} {\bibfnamefont {F.~D.}\ \bibnamefont
  {Gioia}}, \bibinfo {author} {\bibfnamefont {G.}~\bibnamefont {Maniccia}},
  \bibinfo {author} {\bibfnamefont {G.}~\bibnamefont {Montani}},\ and\ \bibinfo
  {author} {\bibfnamefont {J.}~\bibnamefont {Niedda}},\ }\href@noop {}
  {}\bibinfo {howpublished} {Phys. Rev. D \textbf{103}, 103511} (\bibinfo
  {year} {2021})\BibitemShut {NoStop}%
\bibitem [{\citenamefont {Kiefer}\ and\ \citenamefont
  {Singh}(1991)}]{KieferSingh}%
  \BibitemOpen
  \bibfield  {author} {\bibinfo {author} {\bibfnamefont {C.}~\bibnamefont
  {Kiefer}}\ and\ \bibinfo {author} {\bibfnamefont {T.~P.}\ \bibnamefont
  {Singh}},\ }\href@noop {} {}\bibinfo {howpublished} {Phys. Rev. D
  \textbf{44}, 1067} (\bibinfo {year} {1991})\BibitemShut {NoStop}%
\bibitem [{\citenamefont {Kiefer}(2012)}]{KieferBook}%
  \BibitemOpen
  \bibfield  {author} {\bibinfo {author} {\bibfnamefont {C.}~\bibnamefont
  {Kiefer}},\ }\href@noop {} {}\bibinfo {howpublished} {\textit{Quantum
  Gravity}, 3rd ed. (Oxford University Press, Oxford)} (\bibinfo {year}
  {2012})\BibitemShut {NoStop}%
\bibitem [{\citenamefont {Brout}\ and\ \citenamefont
  {Venturi}(1989)}]{BroutVenturi}%
  \BibitemOpen
  \bibfield  {author} {\bibinfo {author} {\bibfnamefont {R.}~\bibnamefont
  {Brout}}\ and\ \bibinfo {author} {\bibfnamefont {G.}~\bibnamefont
  {Venturi}},\ }\href@noop {} {}\bibinfo {howpublished} {Phys. Rev. D
  \textbf{39}, 2436} (\bibinfo {year} {1989})\BibitemShut {NoStop}%
\bibitem [{\citenamefont {Bertoni}\ \emph {et~al.}(1996)\citenamefont
  {Bertoni}, \citenamefont {Finelli},\ and\ \citenamefont {Venturi}}]{Bertoni}%
  \BibitemOpen
  \bibfield  {author} {\bibinfo {author} {\bibfnamefont {C.}~\bibnamefont
  {Bertoni}}, \bibinfo {author} {\bibfnamefont {F.}~\bibnamefont {Finelli}},\
  and\ \bibinfo {author} {\bibfnamefont {G.}~\bibnamefont {Venturi}},\
  }\href@noop {} {}\bibinfo {howpublished} {Classical Quantum Gravity
  \textbf{13}, 2375} (\bibinfo {year} {1996})\BibitemShut {NoStop}%
\bibitem [{\citenamefont {r}\ and\ \citenamefont {Torre}(1991)}]{KucharTorre}%
  \BibitemOpen
  \bibfield  {author} {\bibinfo {author} {\bibfnamefont {K.~V.~K.}\
  \bibnamefont {r}}\ and\ \bibinfo {author} {\bibfnamefont {C.~G.}\
  \bibnamefont {Torre}},\ }\href@noop {} {}\bibinfo {howpublished} {Phys. Rev.
  D \textbf{43}, 419} (\bibinfo {year} {1991})\BibitemShut {NoStop}%
\bibitem [{\citenamefont {Maniccia}\ and\ \citenamefont
  {Montani}(2023)}]{ManicciaKinematical}%
  \BibitemOpen
  \bibfield  {author} {\bibinfo {author} {\bibfnamefont {G.}~\bibnamefont
  {Maniccia}}\ and\ \bibinfo {author} {\bibfnamefont {G.}~\bibnamefont
  {Montani}},\ }\href@noop {} {}\bibinfo {howpublished} {in \textit{The
  Sixteenth Marcel Grossmann Meeting} (World Scientific, Singapore),
  pp.~4146--4158} (\bibinfo {year} {2023})\BibitemShut {NoStop}%
\bibitem [{\citenamefont {Maniccia}\ and\ \citenamefont
  {Montani}(2022)}]{Maniccia2022}%
  \BibitemOpen
  \bibfield  {author} {\bibinfo {author} {\bibfnamefont {G.}~\bibnamefont
  {Maniccia}}\ and\ \bibinfo {author} {\bibfnamefont {G.}~\bibnamefont
  {Montani}},\ }\href@noop {} {}\bibinfo {howpublished} {Phys. Rev. D
  \textbf{105}, 086014} (\bibinfo {year} {2022})\BibitemShut {NoStop}%
\bibitem [{\citenamefont {Maniccia}\ \emph {et~al.}(2024)\citenamefont
  {Maniccia}, \citenamefont {Montani},\ and\ \citenamefont
  {Tosoni}}]{Maniccia2024}%
  \BibitemOpen
  \bibfield  {author} {\bibinfo {author} {\bibfnamefont {G.}~\bibnamefont
  {Maniccia}}, \bibinfo {author} {\bibfnamefont {G.}~\bibnamefont {Montani}},\
  and\ \bibinfo {author} {\bibfnamefont {M.}~\bibnamefont {Tosoni}},\
  }\href@noop {} {}\bibinfo {howpublished} {Phys. Rev. D \textbf{110}, 086012}
  (\bibinfo {year} {2024})\BibitemShut {NoStop}%
\bibitem [{\citenamefont {Brown}\ and\ \citenamefont {r}(1995)}]{BrownKuchar}%
  \BibitemOpen
  \bibfield  {author} {\bibinfo {author} {\bibfnamefont {J.~D.}\ \bibnamefont
  {Brown}}\ and\ \bibinfo {author} {\bibfnamefont {K.~V.~K.}\ \bibnamefont
  {r}},\ }\href@noop {} {}\bibinfo {howpublished} {Phys. Rev. D \textbf{51},
  5600} (\bibinfo {year} {1995})\BibitemShut {NoStop}%
\bibitem [{\citenamefont {Weinberg}(2008)}]{WeinbergCosmology}%
  \BibitemOpen
  \bibfield  {author} {\bibinfo {author} {\bibfnamefont {S.}~\bibnamefont
  {Weinberg}},\ }\href@noop {} {}\bibinfo {howpublished} {\textit{Cosmology}
  (Oxford University Press, Oxford)} (\bibinfo {year} {2008})\BibitemShut
  {NoStop}%
\bibitem [{\citenamefont {Brizuela}\ \emph
  {et~al.}(2016{\natexlab{a}})\citenamefont {Brizuela}, \citenamefont
  {Kiefer},\ and\ \citenamefont {Kr\"amer}}]{BrizuelaDeSitter}%
  \BibitemOpen
  \bibfield  {author} {\bibinfo {author} {\bibfnamefont {D.}~\bibnamefont
  {Brizuela}}, \bibinfo {author} {\bibfnamefont {C.}~\bibnamefont {Kiefer}},\
  and\ \bibinfo {author} {\bibfnamefont {M.}~\bibnamefont {Kr\"amer}},\
  }\href@noop {} {}\bibinfo {howpublished} {Phys. Rev. D \textbf{93}, 104035}
  (\bibinfo {year} {2016}{\natexlab{a}})\BibitemShut {NoStop}%
\bibitem [{\citenamefont {Montani}\ \emph {et~al.}(2011)\citenamefont
  {Montani}, \citenamefont {Battisti}, \citenamefont {Benini},\ and\
  \citenamefont {Imponente}}]{MontaniBook}%
  \BibitemOpen
  \bibfield  {author} {\bibinfo {author} {\bibfnamefont {G.}~\bibnamefont
  {Montani}}, \bibinfo {author} {\bibfnamefont {M.~V.}\ \bibnamefont
  {Battisti}}, \bibinfo {author} {\bibfnamefont {R.}~\bibnamefont {Benini}},\
  and\ \bibinfo {author} {\bibfnamefont {G.}~\bibnamefont {Imponente}},\
  }\href@noop {} {}\bibinfo {howpublished} {\textit{Primordial Cosmology}
  (World Scientific, Singapore)} (\bibinfo {year} {2011})\BibitemShut {NoStop}%
\bibitem [{\citenamefont {Kirillov}\ and\ \citenamefont
  {Montani}(2002)}]{KirillovMontani2002}%
  \BibitemOpen
  \bibfield  {author} {\bibinfo {author} {\bibfnamefont {A.~A.}\ \bibnamefont
  {Kirillov}}\ and\ \bibinfo {author} {\bibfnamefont {G.}~\bibnamefont
  {Montani}},\ }\href@noop {} {}\bibinfo {howpublished} {Phys. Rev. D
  \textbf{66}, 064010} (\bibinfo {year} {2002})\BibitemShut {NoStop}%
\bibitem [{\citenamefont {Montani}(1995)}]{Montani1995}%
  \BibitemOpen
  \bibfield  {author} {\bibinfo {author} {\bibfnamefont {G.}~\bibnamefont
  {Montani}},\ }\href@noop {} {}\bibinfo {howpublished} {On the general
  behaviour of the universe near the cosmological singularity, Classical
  Quantum Gravity \textbf{12}, 2505} (\bibinfo {year} {1995})\BibitemShut
  {NoStop}%
\bibitem [{\citenamefont {Kirillov}\ and\ \citenamefont
  {Montani}(1997)}]{KirillovMontani1997}%
  \BibitemOpen
  \bibfield  {author} {\bibinfo {author} {\bibfnamefont {A.~A.}\ \bibnamefont
  {Kirillov}}\ and\ \bibinfo {author} {\bibfnamefont {G.}~\bibnamefont
  {Montani}},\ }\href@noop {} {}\bibinfo {howpublished} {Origin of a classical
  space in quantum inhomogeneous models, JETP Lett. \textbf{66}, 475} (\bibinfo
  {year} {1997})\BibitemShut {NoStop}%
\bibitem [{\citenamefont {Benini}\ and\ \citenamefont
  {Montani}(2007)}]{BeniniMontani2007}%
  \BibitemOpen
  \bibfield  {author} {\bibinfo {author} {\bibfnamefont {R.}~\bibnamefont
  {Benini}}\ and\ \bibinfo {author} {\bibfnamefont {G.}~\bibnamefont
  {Montani}},\ }\href@noop {} {}\bibinfo {howpublished} {Inhomogeneous quantum
  Mixmaster: From classical towards quantum mechanics, Classical Quantum
  Gravity \textbf{24}, 387} (\bibinfo {year} {2007})\BibitemShut {NoStop}%
\bibitem [{\citenamefont {Doroshkevich}\ \emph {et~al.}(1973)\citenamefont
  {Doroshkevich}, \citenamefont {Lukash},\ and\ \citenamefont
  {Novikov}}]{Doroshkevich1973}%
  \BibitemOpen
  \bibfield  {author} {\bibinfo {author} {\bibfnamefont {A.~G.}\ \bibnamefont
  {Doroshkevich}}, \bibinfo {author} {\bibfnamefont {V.~N.}\ \bibnamefont
  {Lukash}},\ and\ \bibinfo {author} {\bibfnamefont {I.~D.}\ \bibnamefont
  {Novikov}},\ }\href@noop {} {}\bibinfo {howpublished} {Isotropization of
  homogeneous cosmological models, Sov. Phys. JETP \textbf{37}, 739} (\bibinfo
  {year} {1973})\BibitemShut {NoStop}%
\bibitem [{\citenamefont {Belinskii}\ \emph {et~al.}(1970)\citenamefont
  {Belinskii}, \citenamefont {Khalatnikov},\ and\ \citenamefont
  {Lifshitz}}]{BKL1970}%
  \BibitemOpen
  \bibfield  {author} {\bibinfo {author} {\bibfnamefont {V.~A.}\ \bibnamefont
  {Belinskii}}, \bibinfo {author} {\bibfnamefont {I.~M.}\ \bibnamefont
  {Khalatnikov}},\ and\ \bibinfo {author} {\bibfnamefont {E.~M.}\ \bibnamefont
  {Lifshitz}},\ }\href@noop {} {}\bibinfo {howpublished} {Oscillatory approach
  to a singular point in the relativistic cosmology, Adv. Phys. \textbf{19},
  525} (\bibinfo {year} {1970})\BibitemShut {NoStop}%
\bibitem [{\citenamefont {Misner}(1968)}]{Misner1968}%
  \BibitemOpen
  \bibfield  {author} {\bibinfo {author} {\bibfnamefont {C.~W.}\ \bibnamefont
  {Misner}},\ }\href@noop {} {}\bibinfo {howpublished} {The isotropy of the
  universe, Astrophys. J. \textbf{151}, 431} (\bibinfo {year}
  {1968})\BibitemShut {NoStop}%
\bibitem [{\citenamefont {Misner}(1969)}]{MisnerMixmaster}%
  \BibitemOpen
  \bibfield  {author} {\bibinfo {author} {\bibfnamefont {C.~W.}\ \bibnamefont
  {Misner}},\ }\href@noop {} {}\bibinfo {howpublished} {Phys. Rev. Lett.
  \textbf{22}, 1071} (\bibinfo {year} {1969})\BibitemShut {NoStop}%
\bibitem [{\citenamefont {Arnowitt}\ \emph {et~al.}(2008)\citenamefont
  {Arnowitt}, \citenamefont {Deser},\ and\ \citenamefont {Misner}}]{ADM}%
  \BibitemOpen
  \bibfield  {author} {\bibinfo {author} {\bibfnamefont {R.}~\bibnamefont
  {Arnowitt}}, \bibinfo {author} {\bibfnamefont {S.}~\bibnamefont {Deser}},\
  and\ \bibinfo {author} {\bibfnamefont {C.~W.}\ \bibnamefont {Misner}},\
  }\href@noop {} {}\bibinfo {howpublished} {Gen. Relativ. Gravit. \textbf{40},
  1997} (\bibinfo {year} {2008})\BibitemShut {NoStop}%
\bibitem [{\citenamefont {DeWitt}(1967)}]{DeWitt}%
  \BibitemOpen
  \bibfield  {author} {\bibinfo {author} {\bibfnamefont {B.~S.}\ \bibnamefont
  {DeWitt}},\ }\href@noop {} {}\bibinfo {howpublished} {Phys. Rev.
  \textbf{160}, 1113} (\bibinfo {year} {1967})\BibitemShut {NoStop}%
\bibitem [{\citenamefont {Wheeler}(1968)}]{Wheeler}%
  \BibitemOpen
  \bibfield  {author} {\bibinfo {author} {\bibfnamefont {J.~A.}\ \bibnamefont
  {Wheeler}},\ }\href@noop {} {}\bibinfo {howpublished} {in \textit{Battelle
  Rencontres}, edited by C.~M. DeWitt and J.~A. Wheeler (Benjamin, New York),
  p.~242} (\bibinfo {year} {1968})\BibitemShut {NoStop}%
\bibitem [{\citenamefont {Halliwell}(1991)}]{Halliwell}%
  \BibitemOpen
  \bibfield  {author} {\bibinfo {author} {\bibfnamefont {J.~J.}\ \bibnamefont
  {Halliwell}},\ }\href@noop {} {}\bibinfo {howpublished} {in \textit{Quantum
  Cosmology and Baby Universes}, edited by S.~Coleman, J.~B. Hartle, T.~Piran,
  and S.~Weinberg (World Scientific, Singapore), p.~159} (\bibinfo {year}
  {1991})\BibitemShut {NoStop}%
\bibitem [{\citenamefont {Misner}(1972)}]{MisnerMinisuperspace}%
  \BibitemOpen
  \bibfield  {author} {\bibinfo {author} {\bibfnamefont {C.~W.}\ \bibnamefont
  {Misner}},\ }\href@noop {} {}\bibinfo {howpublished} {in \textit{Magic
  Without Magic}, edited by J.~R. Klauder (Freeman, San Francisco), p.~441}
  (\bibinfo {year} {1972})\BibitemShut {NoStop}%
\bibitem [{\citenamefont {Ryan}(1975)}]{RyanShepley}%
  \BibitemOpen
  \bibfield  {author} {\bibinfo {author} {\bibfnamefont {M.~P.}\ \bibnamefont
  {Ryan}},\ }\href@noop {} {}\bibinfo {howpublished} {Jr. and L.~C. Shepley,
  \textit{Homogeneous Relativistic Cosmologies} (Princeton University Press,
  Princeton)} (\bibinfo {year} {1975})\BibitemShut {NoStop}%
\bibitem [{\citenamefont {r}(1992)}]{KucharTime}%
  \BibitemOpen
  \bibfield  {author} {\bibinfo {author} {\bibfnamefont {K.~V.~K.}\
  \bibnamefont {r}},\ }\href@noop {} {}\bibinfo {howpublished} {in
  \textit{Proceedings of the 4th Canadian Conference on General Relativity and
  Relativistic Astrophysics}, edited by G.~Kunstatter, D.~E. Vincent, and J.~G.
  Williams (World Scientific, Singapore), p.~69} (\bibinfo {year}
  {1992})\BibitemShut {NoStop}%
\bibitem [{\citenamefont {Isham}(1993)}]{IshamTime}%
  \BibitemOpen
  \bibfield  {author} {\bibinfo {author} {\bibfnamefont {C.~J.}\ \bibnamefont
  {Isham}},\ }\href@noop {} {}\bibinfo {howpublished} {in \textit{Integrable
  Systems, Quantum Groups, and Quantum Field Theories}, edited by L.~A. Ibort
  and M.~A. Rodr\'iguez (Kluwer, Dordrecht), p.~157} (\bibinfo {year}
  {1993})\BibitemShut {NoStop}%
\bibitem [{\citenamefont {Misner}\ \emph {et~al.}(1973)\citenamefont {Misner},
  \citenamefont {Thorne},\ and\ \citenamefont {Wheeler}}]{MisnerThorneWheeler}%
  \BibitemOpen
  \bibfield  {author} {\bibinfo {author} {\bibfnamefont {C.~W.}\ \bibnamefont
  {Misner}}, \bibinfo {author} {\bibfnamefont {K.~S.}\ \bibnamefont {Thorne}},\
  and\ \bibinfo {author} {\bibfnamefont {J.~A.}\ \bibnamefont {Wheeler}},\
  }\href@noop {} {}\bibinfo {howpublished} {\textit{Gravitation} (W. H.
  Freeman, San Francisco)} (\bibinfo {year} {1973})\BibitemShut {NoStop}%
\bibitem [{\citenamefont {Born}\ and\ \citenamefont
  {Oppenheimer}(1927)}]{BornOppenheimer}%
  \BibitemOpen
  \bibfield  {author} {\bibinfo {author} {\bibfnamefont {M.}~\bibnamefont
  {Born}}\ and\ \bibinfo {author} {\bibfnamefont {R.}~\bibnamefont
  {Oppenheimer}},\ }\href@noop {} {}\bibinfo {howpublished} {Ann. Phys.
  (Leipzig) \textbf{389}, 457} (\bibinfo {year} {1927})\BibitemShut {NoStop}%
\bibitem [{\citenamefont {Bunch}\ and\ \citenamefont
  {Davies}(1978)}]{BunchDavies}%
  \BibitemOpen
  \bibfield  {author} {\bibinfo {author} {\bibfnamefont {T.~S.}\ \bibnamefont
  {Bunch}}\ and\ \bibinfo {author} {\bibfnamefont {P.~C.~W.}\ \bibnamefont
  {Davies}},\ }\href@noop {} {}\bibinfo {howpublished} {Proc. R. Soc. A
  \textbf{360}, 117} (\bibinfo {year} {1978})\BibitemShut {NoStop}%
\bibitem [{\citenamefont {Kiefer}\ and\ \citenamefont
  {Kr\"amer}(2012)}]{KieferKramer}%
  \BibitemOpen
  \bibfield  {author} {\bibinfo {author} {\bibfnamefont {C.}~\bibnamefont
  {Kiefer}}\ and\ \bibinfo {author} {\bibfnamefont {M.}~\bibnamefont
  {Kr\"amer}},\ }\href@noop {} {}\bibinfo {howpublished} {Phys. Rev. Lett.
  \textbf{108}, 021301} (\bibinfo {year} {2012})\BibitemShut {NoStop}%
\bibitem [{\citenamefont {Kamenshchik}\ \emph {et~al.}(2013)\citenamefont
  {Kamenshchik}, \citenamefont {Tronconi},\ and\ \citenamefont
  {Venturi}}]{Kamenshchik}%
  \BibitemOpen
  \bibfield  {author} {\bibinfo {author} {\bibfnamefont {A.~Y.}\ \bibnamefont
  {Kamenshchik}}, \bibinfo {author} {\bibfnamefont {A.}~\bibnamefont
  {Tronconi}},\ and\ \bibinfo {author} {\bibfnamefont {G.}~\bibnamefont
  {Venturi}},\ }\href@noop {} {}\bibinfo {howpublished} {Phys. Lett. B
  \textbf{726}, 518} (\bibinfo {year} {2013})\BibitemShut {NoStop}%
\bibitem [{\citenamefont {Bini}\ \emph {et~al.}(2013)\citenamefont {Bini},
  \citenamefont {Esposito}, \citenamefont {Kiefer}, \citenamefont {Kr\"amer},\
  and\ \citenamefont {Pessina}}]{Bini}%
  \BibitemOpen
  \bibfield  {author} {\bibinfo {author} {\bibfnamefont {D.}~\bibnamefont
  {Bini}}, \bibinfo {author} {\bibfnamefont {G.}~\bibnamefont {Esposito}},
  \bibinfo {author} {\bibfnamefont {C.}~\bibnamefont {Kiefer}}, \bibinfo
  {author} {\bibfnamefont {M.}~\bibnamefont {Kr\"amer}},\ and\ \bibinfo
  {author} {\bibfnamefont {F.}~\bibnamefont {Pessina}},\ }\href@noop {}
  {}\bibinfo {howpublished} {Phys. Rev. D \textbf{87}, 104008} (\bibinfo {year}
  {2013})\BibitemShut {NoStop}%
\bibitem [{\citenamefont {Brizuela}\ \emph
  {et~al.}(2016{\natexlab{b}})\citenamefont {Brizuela}, \citenamefont
  {Kiefer},\ and\ \citenamefont {Kr\"amer}}]{BrizuelaSlowRoll}%
  \BibitemOpen
  \bibfield  {author} {\bibinfo {author} {\bibfnamefont {D.}~\bibnamefont
  {Brizuela}}, \bibinfo {author} {\bibfnamefont {C.}~\bibnamefont {Kiefer}},\
  and\ \bibinfo {author} {\bibfnamefont {M.}~\bibnamefont {Kr\"amer}},\
  }\href@noop {} {}\bibinfo {howpublished} {Phys. Rev. D \textbf{94}, 123527}
  (\bibinfo {year} {2016}{\natexlab{b}})\BibitemShut {NoStop}%
\bibitem [{\citenamefont {Maniccia}\ \emph
  {et~al.}(2023{\natexlab{b}})\citenamefont {Maniccia}, \citenamefont
  {Montani},\ and\ \citenamefont {Torcellini}}]{Maniccia2023}%
  \BibitemOpen
  \bibfield  {author} {\bibinfo {author} {\bibfnamefont {G.}~\bibnamefont
  {Maniccia}}, \bibinfo {author} {\bibfnamefont {G.}~\bibnamefont {Montani}},\
  and\ \bibinfo {author} {\bibfnamefont {L.}~\bibnamefont {Torcellini}},\
  }\href@noop {} {}\bibinfo {howpublished} {Universe \textbf{9}, 169} (\bibinfo
  {year} {2023}{\natexlab{b}})\BibitemShut {NoStop}%
\bibitem [{\citenamefont {Lewis}(1967)}]{Lewis1967}%
  \BibitemOpen
  \bibfield  {author} {\bibinfo {author} {\bibfnamefont {H.~R.}\ \bibnamefont
  {Lewis}},\ }\href@noop {} {}\bibinfo {howpublished} {Jr., J. Math. Phys.
  \textbf{8}, 1976} (\bibinfo {year} {1967})\BibitemShut {NoStop}%
\bibitem [{\citenamefont {Lewis}(1969)}]{LewisRiesenfeld}%
  \BibitemOpen
  \bibfield  {author} {\bibinfo {author} {\bibfnamefont {H.~R.}\ \bibnamefont
  {Lewis}},\ }\href@noop {} {}\bibinfo {howpublished} {Jr. and W.~B.
  Riesenfeld, J. Math. Phys. \textbf{10}, 1458} (\bibinfo {year}
  {1969})\BibitemShut {NoStop}%
\bibitem [{\citenamefont {Pedrosa}(1987)}]{Pedrosa}%
  \BibitemOpen
  \bibfield  {author} {\bibinfo {author} {\bibfnamefont {I.~A.}\ \bibnamefont
  {Pedrosa}},\ }\href@noop {} {}\bibinfo {howpublished} {Phys. Rev. A
  \textbf{36}, 1279} (\bibinfo {year} {1987})\BibitemShut {NoStop}%
\bibitem [{\citenamefont {Dodonov}\ and\ \citenamefont
  {Man'ko}(1989)}]{Dodonov}%
  \BibitemOpen
  \bibfield  {author} {\bibinfo {author} {\bibfnamefont {V.~V.}\ \bibnamefont
  {Dodonov}}\ and\ \bibinfo {author} {\bibfnamefont {V.~I.}\ \bibnamefont
  {Man'ko}},\ }\href@noop {} {}\bibinfo {howpublished} {\textit{Invariants and
  the Evolution of Nonstationary Quantum Systems}, Proc. Lebedev Phys. Inst.
  Vol.~183 (Nova Science, Commack)} (\bibinfo {year} {1989})\BibitemShut
  {NoStop}%
\bibitem [{\citenamefont {Ermakov}(1880)}]{Ermakov}%
  \BibitemOpen
  \bibfield  {author} {\bibinfo {author} {\bibfnamefont {V.~P.}\ \bibnamefont
  {Ermakov}},\ }\href@noop {} {}\bibinfo {howpublished} {Univ. Izv. Kiev
  \textbf{20}, 1} (\bibinfo {year} {1880})\BibitemShut {NoStop}%
\bibitem [{\citenamefont {Pinney}(1950)}]{Pinney}%
  \BibitemOpen
  \bibfield  {author} {\bibinfo {author} {\bibfnamefont {E.}~\bibnamefont
  {Pinney}},\ }\href@noop {} {}\bibinfo {howpublished} {Proc. Am. Math. Soc.
  \textbf{1}, 681} (\bibinfo {year} {1950})\BibitemShut {NoStop}%
\bibitem [{\citenamefont {Grishchuk}\ and\ \citenamefont
  {Sidorov}(1990)}]{GrishchukSidorov}%
  \BibitemOpen
  \bibfield  {author} {\bibinfo {author} {\bibfnamefont {L.~P.}\ \bibnamefont
  {Grishchuk}}\ and\ \bibinfo {author} {\bibfnamefont {Y.~V.}\ \bibnamefont
  {Sidorov}},\ }\href@noop {} {}\bibinfo {howpublished} {Phys. Rev. D
  \textbf{42}, 3413} (\bibinfo {year} {1990})\BibitemShut {NoStop}%
\bibitem [{\citenamefont {\textit{et al.}
  (Planck~Collaboration)}(2020)}]{PlanckNG}%
  \BibitemOpen
  \bibfield  {author} {\bibinfo {author} {\bibfnamefont {Y.~A.}\ \bibnamefont
  {\textit{et al.} (Planck~Collaboration)}},\ }\href@noop {} {}\bibinfo
  {howpublished} {Astron. Astrophys. \textbf{641}, A9} (\bibinfo {year}
  {2020})\BibitemShut {NoStop}%
\bibitem [{\citenamefont {Maldacena}(2003)}]{Maldacena}%
  \BibitemOpen
  \bibfield  {author} {\bibinfo {author} {\bibfnamefont {J.~M.}\ \bibnamefont
  {Maldacena}},\ }\href@noop {} {}\bibinfo {howpublished} {J. High Energy Phys.
  \textbf{05} 013} (\bibinfo {year} {2003})\BibitemShut {NoStop}%
\bibitem [{\citenamefont {Mukhanov}\ \emph {et~al.}(1992)\citenamefont
  {Mukhanov}, \citenamefont {Feldman},\ and\ \citenamefont
  {Brandenberger}}]{MukhanovFeldmanBrandenberger}%
  \BibitemOpen
  \bibfield  {author} {\bibinfo {author} {\bibfnamefont {V.~F.}\ \bibnamefont
  {Mukhanov}}, \bibinfo {author} {\bibfnamefont {H.~A.}\ \bibnamefont
  {Feldman}},\ and\ \bibinfo {author} {\bibfnamefont {R.~H.}\ \bibnamefont
  {Brandenberger}},\ }\href@noop {} {}\bibinfo {howpublished} {Phys. Rep.
  \textbf{215}, 203} (\bibinfo {year} {1992})\BibitemShut {NoStop}%
\bibitem [{\citenamefont {Baumann}(2011)}]{Baumann}%
  \BibitemOpen
  \bibfield  {author} {\bibinfo {author} {\bibfnamefont {D.}~\bibnamefont
  {Baumann}},\ }\href@noop {} {}\bibinfo {howpublished} {in \textit{Physics of
  the Large and the Small}, edited by C.~Cs\'aki and S.~Dodelson (World
  Scientific, Singapore), p.~523} (\bibinfo {year} {2011})\BibitemShut
  {NoStop}%
\bibitem [{\citenamefont {Fr\"ohlich}(1952)}]{Frohlich}%
  \BibitemOpen
  \bibfield  {author} {\bibinfo {author} {\bibfnamefont {H.}~\bibnamefont
  {Fr\"ohlich}},\ }\href@noop {} {}\bibinfo {howpublished} {Proc. R. Soc. A
  \textbf{215}, 291} (\bibinfo {year} {1952})\BibitemShut {NoStop}%
\bibitem [{\citenamefont {Bardeen}\ \emph {et~al.}(1957)\citenamefont
  {Bardeen}, \citenamefont {Cooper},\ and\ \citenamefont {Schrieffer}}]{BCS}%
  \BibitemOpen
  \bibfield  {author} {\bibinfo {author} {\bibfnamefont {J.}~\bibnamefont
  {Bardeen}}, \bibinfo {author} {\bibfnamefont {L.~N.}\ \bibnamefont
  {Cooper}},\ and\ \bibinfo {author} {\bibfnamefont {J.~R.}\ \bibnamefont
  {Schrieffer}},\ }\href@noop {} {}\bibinfo {howpublished} {Phys. Rev.
  \textbf{108}, 1175} (\bibinfo {year} {1957})\BibitemShut {NoStop}%
\bibitem [{\citenamefont {Mahan}(2000)}]{Mahan}%
  \BibitemOpen
  \bibfield  {author} {\bibinfo {author} {\bibfnamefont {G.~D.}\ \bibnamefont
  {Mahan}},\ }\href@noop {} {}\bibinfo {howpublished} {\textit{Many-Particle
  Physics}, 3rd ed. (Kluwer Academic/Plenum, New York)} (\bibinfo {year}
  {2000})\BibitemShut {NoStop}%
\bibitem [{\citenamefont {Wald}(1983)}]{WaldNoHair}%
  \BibitemOpen
  \bibfield  {author} {\bibinfo {author} {\bibfnamefont {R.~M.}\ \bibnamefont
  {Wald}},\ }\href@noop {} {}\bibinfo {howpublished} {Phys. Rev. D \textbf{28},
  2118} (\bibinfo {year} {1983})\BibitemShut {NoStop}%
\end{thebibliography}%

\end{document}